\documentclass[11pt,a4paper]{article}
\usepackage{jheppub}
\usepackage{amsmath,amssymb}
\usepackage{slashed}
\usepackage{graphicx}%
\usepackage{subfigure}
\usepackage{mathptmx}
\usepackage{url}
\newcommand{\beq}{\begin{equation}}
\newcommand{\eeq}{\end{equation}}

\newcommand{\be}{\begin{eqnarray}}
\newcommand{\ee}{\end{eqnarray}}
\long\def\hidestart#1\hideend{}
\title
{Exploring autocorrelations in two-flavour Wilson Lattice QCD using 
DD-HMC algorithm}

\author{Abhishek Chowdhury$^{a}$,}
\author{Asit K. De$^{a}$,}
\author{Sangita De Sarkar$^{a}$,}
\author{A. Harindranath$^{a}$\footnote{Corresponding Author},}
\author{Jyotirmoy Maiti$^{b}$,}
\author{Santanu Mondal$^{a}$ and}
\author{Anwesa Sarkar$^{a}$}

\affiliation{$^{a}$Theory Division, Saha Institute of Nuclear Physics \\
 1/AF Bidhan Nagar, Kolkata 700064, India}

\affiliation{$^{b}$Department of Physics, Barasat Government College,\\
10 KNC Road, Barasat, Kolkata 700124, India}

\emailAdd{abhishek.chowdhury@saha.ac.in}
\emailAdd{asitk.de@saha.ac.in}
\emailAdd{sangita.desarkar@saha.ac.in}
\emailAdd{a.harindranath@saha.ac.in}
\emailAdd{jyotirmoy.maiti@gmail.com}
\emailAdd{santanu.mondal@saha.ac.in}
\emailAdd{anwesa.sarkar@saha.ac.in}

\date{August September 12, 2012}
\abstract{
We perform an extensive study of autocorrelation of several observables in 
lattice QCD with two degenerate 
flavours of naive Wilson fermions and unimproved
Wilson gauge action using DD-HMC algorithm. We show that 
(1) at a given lattice spacing, autocorrelation of topological
susceptibility  
 decreases with decreasing quark mass and autocorrelations of plaquette and Wilson loop
do not increase with decreasing quark mass,
(2) autocorrelation of topological susceptibility 
substantially increases with decreasing lattice spacing
but autocorrelation of 
topological charge density correlator shows only mild increase and
(3) increasing the size and the smearing level increase the
autocorrelation of Wilson loop.}

\begin{document}
\maketitle
%%%%%%%%%%%%%%%%%%%%%%%%%%%%%%%%%%%%%%%%%%%%%%%%%%%%%%%%%%
\section{Introduction}\label{intro}
%%%%%%%%%%%%%%%%%%%%%%%%%%%%%%%%%%%%%%%%%%%%%%%%%%%%%%%%%%
The most popular algorithm to simulate lattice QCD with Dynamical 
fermions is the Hybrid Monte Carlo (HMC) \cite{hmc} and one of its improved 
variations, namely, Domain Decomposed Hybrid Monte Carlo (DD-HMC) 
\cite{ddhmc} aims to achieve significant acceleration of the numerical 
simulation. Dynamical Wilson fermion simulations  at smaller quark masses,
smaller lattice spacings and 
larger lattice volumes on currently available computers have become feasible 
with recent developments such as DD-HMC algorithm. However, approach to the 
continuum and chiral limits may still be hampered by the phenomenon 
of critical slowing down. One of the manifestation of critical slowing down 
is the increase in autocorrelation times associated with the measurements of 
various observables. Thus measurements of autocorrelation times 
help us to evaluate the performance of an algorithm in terms of  
critical slowing down.      
In addition, an accurate determination of the uncertainty associated with 
the measurement of an observable requires a realistic estimation of 
the autocorrelation of the observable which in turn depends on the
various parameters associated with the particular algorithm used.

An extensive study of autocorrelation mainly in pure $SU(3)$ 
gauge theory using DD-HMC algorithm 
has been carried out by
ALPHA collaboration \cite{sommer}. They have shown that 
the autocorrelation of squared topological charge increases dramatically with decreasing lattice spacing
while Wilson loops decouple from the modes which slow down the topological charge as lattice spacing
decreases. In the simulations with dynamical fermions, the study becomes more difficult, 
because the autocorrelation may now depend on number of quark 
flavours ($n_f$), the quark masses and the fermion action used 
\cite{luscher}. In fact ALPHA collaboration \cite{sommer} has shown, in the case of
$n_f=2$
QCD with Clover action for a given value of quark mass and lattice volume, that squared
topological charge decorrelates faster compared with pure gauge at
approximately same lattice spacing.
These dependencies and the one on the lattice spacing remain to be studied in detail.    
In this work we study the 
autocorrelations of a variety of observables measured with DD-HMC 
algorithm in the case of naive Wilson fermions \cite{wilson1, wilson2}.    
Note that  the measurement of autocorrelation is notoriously difficult, since 
accurate 
determination of it  may require considerably larger accumulated
statistics (total molecular dyanamics time).
In this work we mainly focus on various trends of autocorrelations that 
we can observe clearly rather than the precise measurement of the integrated autocorrelation time.

%%%%%%%%%%%%%%%%%%%%%%%%%%%%%%%%%%%%%%%%%%%%%%%%%%%%%%%%%%%%%%%%%%%%
\section{Autocorrelation}
%%%%%%%%%%%%%%%%%%%%%%%%%%%%%%%%%%%%%%%%%%%%%%%%%%%%%%%%%%%%%%%%%%%%
%\subsection{Algorithm}

%\subsection{Observables}

Following Refs. \cite{sommer} and \cite{sokal} let us assume that ${\cal O}=\Big\{{\cal O}(x)\Big\}_{x \in S}$ 
be a real-valued function defined on the state space $S$ that is 
square integrable with respect to $\pi$, where $\pi$
is the stationary Markov chain probability distribution with probability 
transition matrix $P$. Now consider that the Markov chain is 
in equilibrium. Then the unnormalized autocorrelation function,   
\begin{eqnarray}
C^{\mathcal{O}}\left(t\right)&=& \langle\mathcal{O}(s) 
\mathcal{O}(s+t)\rangle- \mu_{\mathcal{O}}^2\nonumber\\
&=&\sum_{x,y}{\mathcal O}(x)\left[\pi_xP_{xy}^{
(\mid t \mid)}-\pi_x\pi_y \right]{\mathcal O}(y)\label{autocorr_func} 
\end{eqnarray}
where $\mu_{\mathcal{O}} \equiv \langle \mathcal{O}(t) 
\rangle = \sum_x \pi_x \mathcal{O}(x)$.
Now if the algorithm satisfies detailed balance, i.e.
$\pi_x P_{xy}= \pi_y P_{yx}$ for all $x,y \in S$ then 
it is convenient to introduce the symmetric matrix
\begin{eqnarray} 
T_{x,y}=\pi_x^{\frac{1}{2}}P_{xy}\pi_y^{-\frac{1}{2}} 
\end{eqnarray}
which has real eigenvalues $\lambda_n$, $n\geq 0$ 
with $\lambda_0=1$ and $\mid \lambda_n \mid < 1$ for $n \geq 1$,
assuming an ergodic algorithm. We order the eigenvalues as 
$\mid \lambda_n \mid \leq \mid \lambda_{n-1}\mid$. 
There is a complete set of eigenfunctions $\chi_n(x)$ 
with $\chi_0(x)=\pi_x^{\frac{1}{2}}$.
By using spectral representation of $T$, Eq. (\ref{autocorr_func}) 
can be reduced to 
\begin{eqnarray}
C^{\mathcal{O}}\left(t\right)=\sum_{n \geq 1}(\lambda_n)^t~ 
\mid \eta_n(\mathcal {O})\mid^2
\end{eqnarray}
where $\eta_n(\mathcal {O})=\sum_x 
\mathcal {O}(x)\chi_n(x)\pi_x^{\frac{1}{2}}$.
Since $\mid \lambda_n \mid < 1$ for $n \geq 1$ 
\begin{eqnarray}
C^{\mathcal{O}}\left(t\right)=\sum_{n 
\geq 1}e^{-t/\tau_n} \mid \eta_n(\mathcal {O})\mid^2
\end{eqnarray} 
where $\tau_n=-\frac{1}{\ln\lambda_n}$, assuming $\lambda_n$'s are positive.

For any particular observable ${\mathcal{O}}$, autocorrelation among the
generated configurations are generally determined by the integrated
autocorrelation time $\tau_{\rm int}^{\mathcal{O}}$ for that
observable. For this purpose, at first, one needs to calculate the
unnormalized autocorrelation function of the observable ${\mathcal{O}}$
measured on a sequence of $N$ equilibrated configurations as  
\begin{equation}
C^{\mathcal{O}}\left(t\right) = \frac{1}{N - t}\sum_{r = 1}^{N - t}
\left({\mathcal{O}}_r - {\overline{\mathcal{O}}} \right)
\left({\mathcal{O}}_{r+t} - {\overline{\mathcal{O}}} \right)~
\end{equation}
where ${\overline{\mathcal{O}}}=\frac{1}{N}\sum_{r=1}^N \mathcal{O}_r $ is
the ensemble average.
Following the {\it windowing} method as recommended by Ref.
 \cite{sokal}, the integrated autocorrelation time is defined as
\begin{equation}
\tau_{\rm int}^{\mathcal{O}} = \frac{1}{2} + \sum_{t=1}^{\rm W}
\Gamma^{\mathcal{O}}\left(t\right)
\label{taudef}
\end{equation}
where  $\Gamma^{\mathcal{O}}\left(t\right) = C^{\mathcal{O}}\left(t\right)/
C^{\mathcal{O}}\left(0\right) $ is the normalized autocorrelation function and 
$W$ is the summation window. 
To calculate the errors, we follow the standard techniques available in the literature 
\cite{anderson, priestley, sokal, wolff, luscher}. 
The variance of $\Gamma^{\mathcal{O}}\left(t\right)$ is given by
\begin{eqnarray}
\langle (\delta\Gamma^{\mathcal{O}}\left(t\right))^2 \rangle\approx \frac{1}{N}\sum_{k=1}^{\infty}[\Gamma^{\mathcal{O}}\left(k+t\right)
+\Gamma^{\mathcal{O}}\left(k-t\right)-2\Gamma^{\mathcal{O}}\left(t\right)\Gamma^{\mathcal{O}}\left(k\right)]^2\label{fn-err}
\end{eqnarray}  
and the variance of $\tau_{int}^{\mathcal{O}}$,
\begin{eqnarray}
\langle (\delta\tau_{int}^{\mathcal{O}} )^2 \rangle\approx \frac{2(2W+1)}{N}(\tau_{int}^{\mathcal{O}})^2.\label{tau-err}
\end{eqnarray}
Different strategies have been suggested in the literature \cite{sokal, wolff, luscher} for choosing
$W$. 
We choose 
$W$ where error of $\Gamma^{\mathcal{O}}\left(t\right)$ becomes equal to
$\Gamma^{\mathcal{O}}\left(t\right)$ \cite{luscher}.
The above expressions are used to calculate the errors unless otherwise stated.
% when enough ($\geq 1000$) statistics is available for measurements.
%Otherwise the errors are calculated
%by the single omission jackknife method. 
In case the total accumulated statistics is
 extremely large an alternative procedure may be to use 
 binning, with binsizes much larger than $\tau_{int}$ for calculating the error \cite{basak}.
% one can use binning, with bins of the order of $\tau_{int}$ 
%for calculating the error \cite{basak}, which we do not 
%pursue in this work.
%%%%%%%%%%%%%%%%%%%%%%%%%%%%%%%%%%%%%%%%%%%%%%%%%%%%
\section{Observables}
%%%%%%%%%%%%%%%%%%%%%%%%%%%%%%%%%%%%%%%%%%%%%%%%%%%%

Let us denote plaquette and Wilson loop of size $R\times T$ 
with smear level $s$
by $P_s$ and $W_s(R,T)$ respectively. Topological susceptibility 
with smear level $s$
is denoted by $Q_s^2$ (the normalization factor, inverse of lattice 
volume, is ignored). 
We have measured the autocorrelations  for the 
plaquette, Wilson loop,  nucleon propagator, pion propagator,  
topological susceptibility 
and topological charge density correlator 
($C(r)=\langle q(x)q(0) \rangle$ where
$q(x)$ is topological charge density and $r=\mid x\mid$)
for the saved 
configurations except for the unsmeared plaquette where we have measured for all the configurations,
at two values of gauge coupling ($\beta = 5.6$ and  $5.8$)
and several values of the hopping parameter $\kappa$.

For pion and nucleon 
we consider the following zero spatial momentum correlation
functions
\begin{eqnarray}
& & C_1(t)~~ =~~\langle 0 \mid {\cal O}^\dagger(t) {\cal O}(0)
\mid 0 \rangle~ ~~{\rm and }~~
 C_2(t)~~ =~~\langle 0 \mid {\cal O}_1^\dagger(t) {\cal O}_2(0) \mid
0 \rangle ~~ 
\end{eqnarray}
where $t$ refers to Euclidean time.
For the nucleon 
${\cal O^\dagger O}\equiv N^\dagger N$ with $N= (q_d^T C \gamma_5 q_u) q_u$.  
For the pion ${\cal O^\dagger O}\equiv P^\dagger P$ or $A^\dagger A$ 
and ${({\cal O}_1)^\dagger{\cal O}_2}= 
A^\dagger P$ or $P^\dagger A$
with  $P = {\overline q}_i \gamma_5 q_j~~~ {\rm and}~~ 
A_4 = {\overline q}_i \gamma_4 \gamma_5 q_j~$  denote the pseudoscalar
density and fourth component of the axial vector current ($i$ and $j$ stand for
flavor indices for the $u$ and $d$ quarks and  for the charged pion $i\ne
j$). For both pion and nucleon we use wall source and point sink.
We measure the autocorrelation of the zero spatial momentum correlation 
functions at an appropriate
time slice corresponding to the plateau region of the 
effective mass. For lattice volume $24^3\times 48$ and $32^3\times 64$ we use $12^{th}$ and
$15^{th}$ time slices respectively. 
 For topological charge density, we use the lattice 
approximation developed for 
$SU(2)$ by DeGrand, Hasenfratz and Kovacs \cite{degrand}, modified for
$SU(3)$ by Hasenfratz and Nieter \cite{hasenfratz1} and implemented in
the MILC code \cite{milc}.
To suppress the ultraviolet lattice artifacts, smearing of link fields 
is employed.  Unless otherwise stated  $20$ HYP smearing steps
with optimized smearing coefficients $\alpha =0.75$,
$\alpha_2=0.6$ and $\alpha_3=0.3$ \cite{hasenfratz2} are used for the gauge observables.
For observables with hadronic operators no gauge field smearing has been used. 
Our data for topological charge, susceptibility and charge density correlator 
are presented in \cite{topo,latt2011,topo-corr}
%**********************************************************************%
\begin{table}
\begin{center}
\begin{tabular}{|l|l|l|l|l|l|l|l|l|l|}

 \multicolumn{7}{c}{$\beta = 5.6$} \\
%&\multicolumn{2}{c|}{$am_\rho$}&
%\multicolumn{1}{c|}{$aF^{V}_\rho$}\\
%\cline{2-4}
\hline
$tag$&$lattice$& $\kappa$& $block $&{$N_2$}& {$N_{trj}$}  &{$\tau$}\\ 
\hline
%{$16^3\times32$}&{$0.156$}&{HMC}&{5000}&{200}&{0.5}&{16(6)}&{28(10)(200)}&{21(7)}&{10(5)(200)}&{} \\
%{$~~~~~,,$}&{$0.157$}&{HMC}&{5000}&{200}&{0.5}&{23(8)}&{27(6)(200)}&{32(8)}&{15(5)(200)}&{} \\
%{$~~~~~,,$}&{$0.1575$}&{HMC}&{5000}&{200}&{0.5}&{26(7)}&{26(7)(200)}&{34(9)}&{16(5)(200)}&{} \\ 
%{$~~~~~,,$}&{$0.158$}&{HMC}&{5000}&{200}&{0.5}&{25(9)}&{36(10)(200)}&{83(20)}&{29(9)(200)}&{} \\ 
%{$A_{1a}$}&{$16^3\times32$}&{$0.15775$}&{$~~~~8^4$}&{$8$}&{$6400$}&{$0.5$} \\ 
%{$A_{1b}$}&{$~~~~~,,$}&{$0.15775$}&{$8^3\times16$}&{$8$}&{$3200$}&{$0.5$} \\
%\hline 
%**********************************************************************
%\hline
%{$B_{1a}$}&{$24^3\times48$}&{$0.1575$}&{$6^3\times8$}&{$8$}&{$9360$}&{$0.5$} \\
{$B_{1b}$}&{$24^3\times48$}&{$0.1575$}&{$12^2\times6^2$}&{$18$}&{$13128$}&{$0.5$} \\
%{$B_{2a}$}&{$~~~~~,,$}&{$0.15775$}&{$6^3\times8$}&{$8$}&{$10560$}&{$0.5$} \\
%{$B_{2b}$}&{$~~~~~,,$}&{$0.15775$}&{$12^2\times6^2$}&{$8$}&{4992}&{0.5}\\
%{$B_{2c}$}&{$~~~~~,,$}&{$0.15775$}&{$12^3\times6$}&{$18$}&{$4448$}&{$0.5$}\\
{$B_{3a}$}&{$~~~~~,,$}&{$0.158$}&{$6^3\times8$}&{$6$}&{$7200$}&{$0.5$} \\
{$B_{3b}$}&{$~~~~~,,$}&{$0.158$}&{$12^2\times6^2$}&{$18$}&{$13646$}&{$0.5$} \\
{$B_{4a}$}&{$~~~~~,,$}&{$0.158125$}&{$6^3\times8$}&{$8$}&{$9360$}&{$0.5$}\\
{$B_{4b}$}&{$~~~~~,,$}&{$0.158125$}&{$12^2\times6^2$}&{$18$}&{$11328$}&{$0.5$}\\
%{$~~~~~,,$}&{$0.158125$}&{$6^3\times8$}&{11808}&{492}&{0.5}&{62(18)}&{122(35)(123)}&{635(69)}&{327(42)(492)}&{}\\
{$B_{5a}$}&{$~~~~~,,$}&{$0.15825$}&{$6^3\times8$}&{$8$}&{$6960$}&{$0.5$}\\
{$B_{5b}$}&{$~~~~~,,$}&{$0.15825$}&{$12^2\times6^2$}&{$18$}&{$12820$}&{$0.5$}\\
%{$B_{5c}$}&{$~~~~~,,$}&{$0.15825$}&{$12^2\times6^2$}&{$14$}&{$4224$}&{$0.5$}\\
%{$B_{5d}$}&{$~~~~~,,$}&{$0.15825$}&{$12^2\times6^2$}&{$22$}&{$3156$}&{$0.5$}\\
%{$~~~~~,,$}&{$0.15825$}&{$6^3\times8$}&{9360}&{390}&{0.5}&{()}&{()(90)}&{238(30)}&{161(22)(390)}&{107(18)} \\
%{$~~~~~,,$}&{$0.15825$}&{$6^3\times8$}&{8640}&{360}&{0.5}&{62(18)}&{133(50)}&{()}&{()}&{}\\
%**********************************************************************
\hline
%{$32^3\times64$}&{$0.15775$}&{$8^3\times16$}&{6080}&{190}&{0.5}&{86(40)}&{443(93)(94)}&{723(134)}&{416(152)(94)}&{}\\
{$C_1$}&{$32^3\times64$}&{$0.15775$}&{$8^3\times16$}&{$8$}&{$6844$}&{$0.5$}\\
{$C_2$}&{$~~~~~,,$}&{$0.158$}&{$8^3\times16$}&{$8$}&{$7576$}&{$0.5$}\\
%{$~~~~~,,$}&{$0.15815$}&{$8^3\times16$}&{11520}&{360}&{0.5}&{115(41)}&{210(84)(90)}&{}&{()}&{}\\
{$C_3$}&{$~~~~~,,$}&{$0.15815$}&{$8^3\times16$}&{$8$}&{$9556$}&{$0.5$}\\
{$C_4$}&{$~~~~~,,$}&{$0.15825$}&{$8^3\times16$}&{$8$}&{$4992$}&{$0.25$}\\
{$C_5$}&{$~~~~~,,$}&{$0.1583$}&{$8^3\times16$}&{$8$}&{$13232$}&{$0.25$}\\
%{$~~~~~,,$}&{$0.1584$}&{$8^3\times16$}&{3904}&{122}&{0.25}&{37(25)}&{25(18)}&{338(112)}&{127(54)}&{} \\
%{$C_6$}&{$~~~~~,,$}&{$0.1584$}&{$8^3\times16$}&{$8$}&{15488}&{0.25}\\
\hline \hline

%***********************************************************************
%***********************************************************************
  \multicolumn{7}{c}{$\beta = 5.8$} \\
\hline
$tag$&$lattice$& $\kappa$& $block$ &{$N_2$}& {$N_{trj}$}  &{$\tau$} \\
\hline
%{$32^3\times64$}&{$0.154$}&{$8^3\times16$}&{6400}&{200}&{0.5}&{20(8)}&{(200)}&{74(23)}&{28(18)(rej=500)(200)}&{} \\
%{$~~~~~,,$}&{$0.1541$}&{$8^3\times16$}&{6400}&{200}&{0.5}&{69(27)}&{(100)}&{}&{(100)}&{} \\
{$D_1$}&{$32^3\times64$}&{$0.1543$}&{$8^3\times16$}&{$8$}&{$9600$}&{$0.5$}\\
%{$D_2$}&{$~~~~~,,$}&{$0.15455$}&{$8^3\times16$}&{$8$}&{$12160$}&{$0.5$}\\
%{$D_3$}&{$~~~~~,,$}&{$0.15462$}&{$8^3\times16$}&{$8$}&{$6240$}&{$0.5$} \\
%{$D_4$}&{$~~~~~,,$}&{$0.1547$}&{$8^3\times16$}&{$8$}&{$10508$}&{$0.375$}\\
{$D_5$}&{$~~~~~,,$}&{$0.15475$}&{$8^3\times16$}&{$8$}&{$6820$}&{$0.25$}\\
\hline \hline
\end{tabular}
\end{center}
%\end{center}
\caption{Lattice parameters and simulation statistics.
Here $block$, $N_2$, $N_{trj}$, $\tau$
refers to DD-HMC block, step number for the force $F_2$, number of 
DD-HMC trajectories and the Molecular Dynamics trajectory length respectively.   }
\label{table1}
\end{table}
 \begin{figure}
\subfigure{
\includegraphics[width=2.8in,clip]
{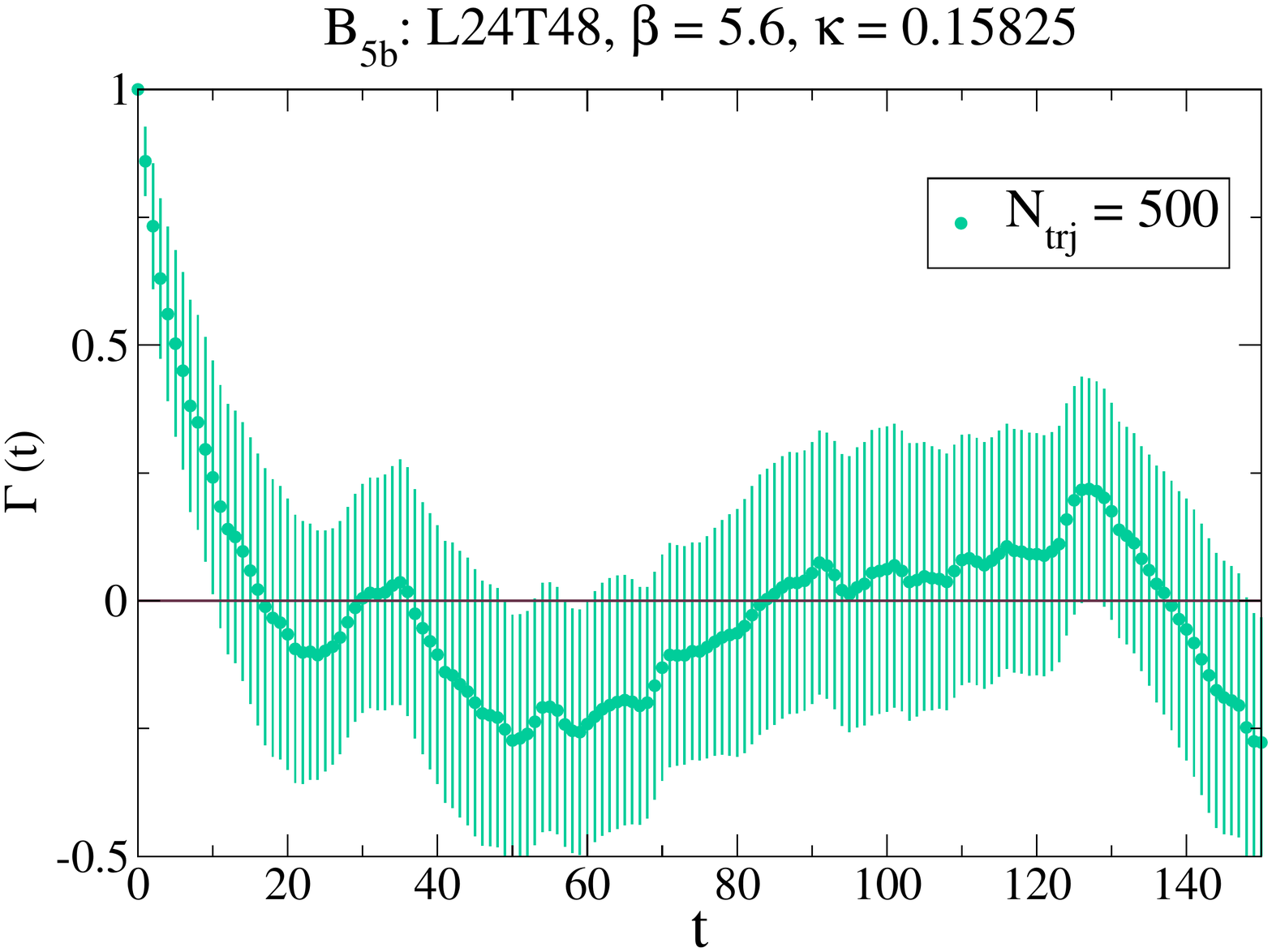}}
\subfigure{
\includegraphics[width=2.8in,clip]
{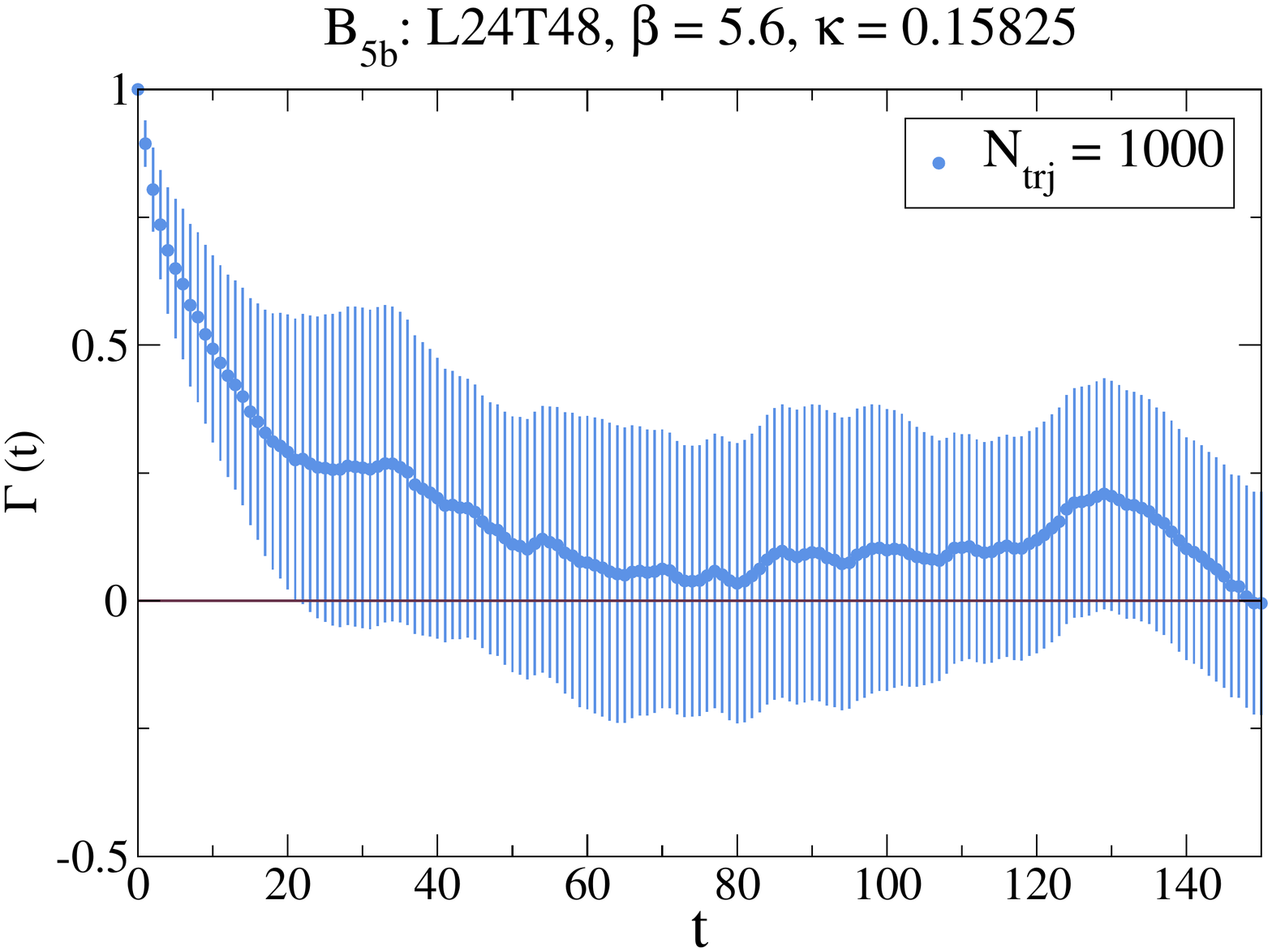}}

\center{\subfigure{
\includegraphics[width=2.8in,clip]
{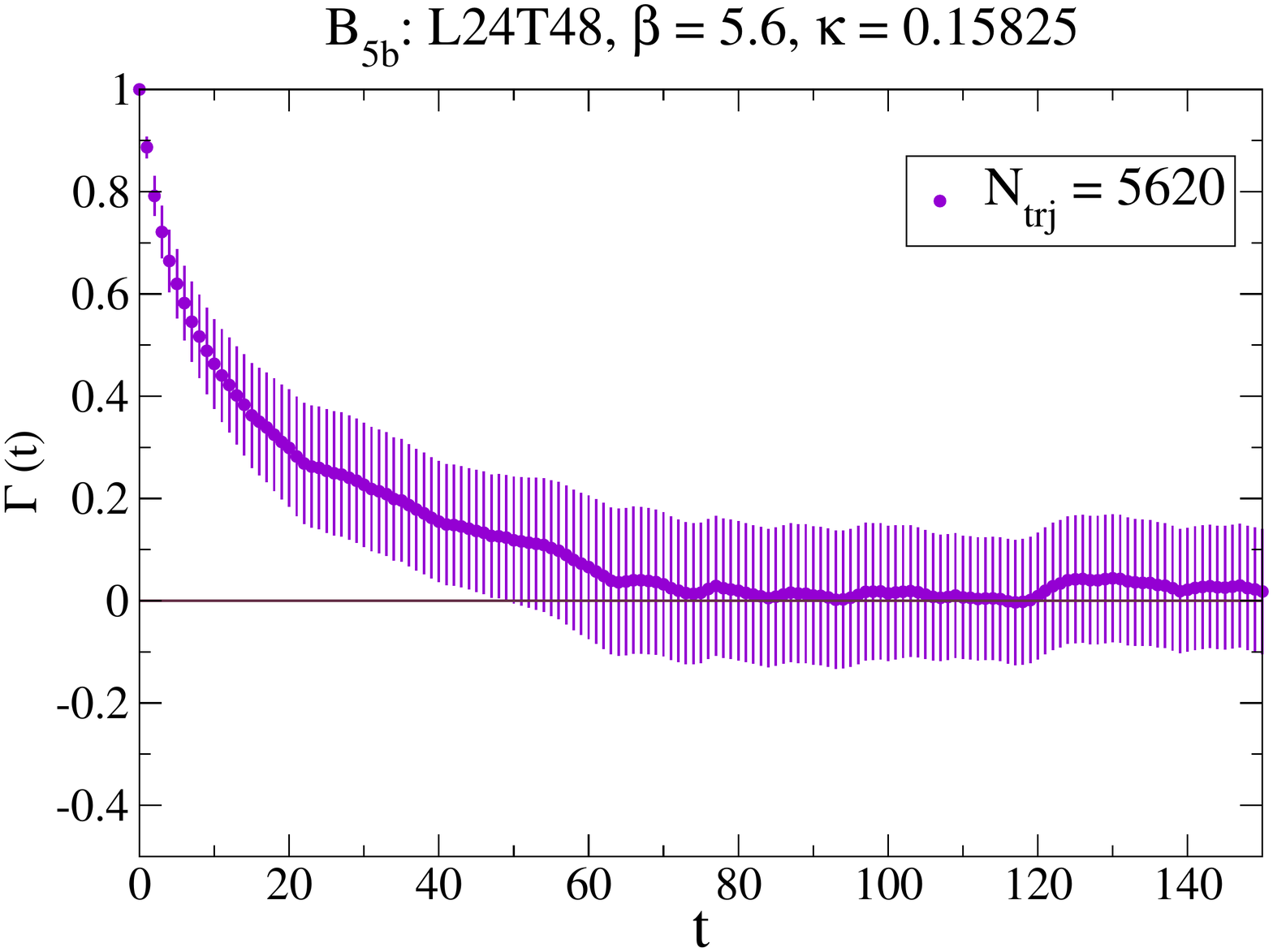}}}
\caption{Autocorrelation functions for the unsmeared plaquette for the total
accumulated statistics 500, 1000 and 5620 respectively at $\beta=5.6$ for the ensemble $B_{5b}$.}
\label{fig_traj_plaq}
\end{figure}

% \begin{figure}
%\subfigure{
%\includegraphics[width=2.8in,clip]
%{autocorr_length_plaq_48x24x24x24p64b56k15825c00000id1_11_N2_18_1gap_500cg.pdf}}
%\subfigure{
%\includegraphics[width=2.8in,clip]
%{autocorr_length_plaq_48x24x24x24p64b56k15825c00000id1_11_N2_18_1gap_1000cg.pdf}}

%\center{\subfigure{
%\includegraphics[width=2.8in,clip]
%{autocorr_length_plaq_48x24x24x24p64b56k15825c00000id1_11_N2_18_1gap_5620cg.pdf}}}
%\caption{Autocorrelation functions for the unsmeared plaquette for the total
%accumulated statistics 500, 1000 and 5620 respectively for $\beta$=5.6, $\kappa=0.15825$ and 
%lattice volume $24^3 \times 48$. Errors are calculated using single omission jackknife technique.}
%\label{fig_traj_plaq1}
%\end{figure}
%%%%%%%%%%%%%%%%%%%%%%%%%%%%%%%%%%%%%%%%%%%%%%%%%%%%%%%%%%%%%%%%%%
\section{Auto-correlation Measurements}
%%%%%%%%%%%%%%%%%%%%%%%%%%%%%%%%%%%%%%%%%%%%%%%%%%%%%%%%%%%%%%%%%%

We have generated ensembles of gauge configurations by means of   
DD-HMC \cite{ddhmc}
algorithm using unimproved Wilson fermion and gauge 
actions \cite{wilson1,wilson2} with 
$n_f=2$ mass degenerate quark flavors. At $\beta=5.6$ the lattice volumes are 
 $24^3 \times 48$ and $32^3 \times 64$ and the renormalized 
physical quark mass (calculated using axial Ward identity)
ranges between $25$ to $125$ MeV ($\overline{\rm MS}$ scheme 
 at $2$ GeV). At $\beta = 5.8$ the lattice 
volume is $32^3 \times 64$ and the renormalized physical quark mass 
 ranges 
from $15$ to $75$ MeV.
To determine the lattice spacing we plot the ratio of lattice 
pion mass to lattice nucleon mass versus lattice pion mass. Extrapolation of the ratio
to the physical point gives the lattice spacing $a$. 
%By plotting the same ratio versus $r_0m_{\pi}$ ($r_0$ is the Sommer parameter)
%we also determine $r_0$ at the physical point.
%We use Sommer parameter method as a second determination of lattice spacing where
%using the value of $r_0$ obtained in hadron mass ratio method we extract the lattice spacing
%$a$. We have found both methods give the same lattice spacing.   
%The lattice spacings are determined using 
%nucleon mass to pion mass ratio and
%Sommer method. 
The lattice spacings at
$\beta =5.6$ and $5.8$ are $0.069$ and $0.053$ fm respectively.
The Sommer method of  determining the scale
agree with the quoted value 
of lattice spacings at $\beta = 5.6 $ and $\beta = 5.8$
for the value of Sommer parameter 
$r_0= 0.44$ fm.    
 
%The configurations for lattice volumes $24^3 \times 48$ and $32^3 \times 64$ 
%and for $16^3 \times 32$ at $\kappa = 0.15775$ are generated using DD-HMC
%algorithm.
The number of thermalized configurations ranges from $7000$ to $14000$.
The lattice parameters and simulation statistics are given in Table 
\ref{table1}. For all ensembles of configurations the average 
Metropolis acceptance rates range between $75-98\%$.

%\input table_lattice_info_nucleon_plus_pion_final.tex
%\input latt2011_table.tex
%%%%%%%%%%%%%%%%%%%%%%%%%%%%%%%%%%%%%%%%%%%%%%%%%%%%%%%%%%%%%%%%%%
\section{Results}
%%%%%%%%%%%%%%%%%%%%%%%%%%%%%%%%%%%%%%%%%%%%%%%%%%%%%%%%%%%%%%%%%
In Fig. \ref{fig_traj_plaq} we show the  
autocorrelation function for the unsmeared plaquette for the total
accumulated statistics 
500, 1000 and 5620 respectively for $\beta$=5.6, $\kappa=0.15825$ and 
lattice volume $24^3 \times 48$. We notice that for smaller statistics, the 
autocorrelation function touches zero earlier
leading to the underestimation of $\tau_{int}$.
Also the positivity of the autocorrelation function is violated in 
contrast to theoretical expectations but the situation improves as 
statistics increases. 
%The errors are calculated by using Eq. (\ref{fn-err}).
%In Fig. \ref{fig_traj_plaq1} we present the same data where errors are calculated
%by using single omission jackknife technique. 
%The observations remain the same as in Fig. \ref{fig_traj_plaq}.
%Note that different symbols for errors are used in Fig. \ref{fig_traj_plaq} and Fig. \ref{fig_traj_plaq1} to
%distinguish between two methods of calculating the
%errors and the same convention is used in the rest of the paper.   
%*************************************************************%
\begin{figure}
\subfigure{
\includegraphics[width=2.8in,clip]
{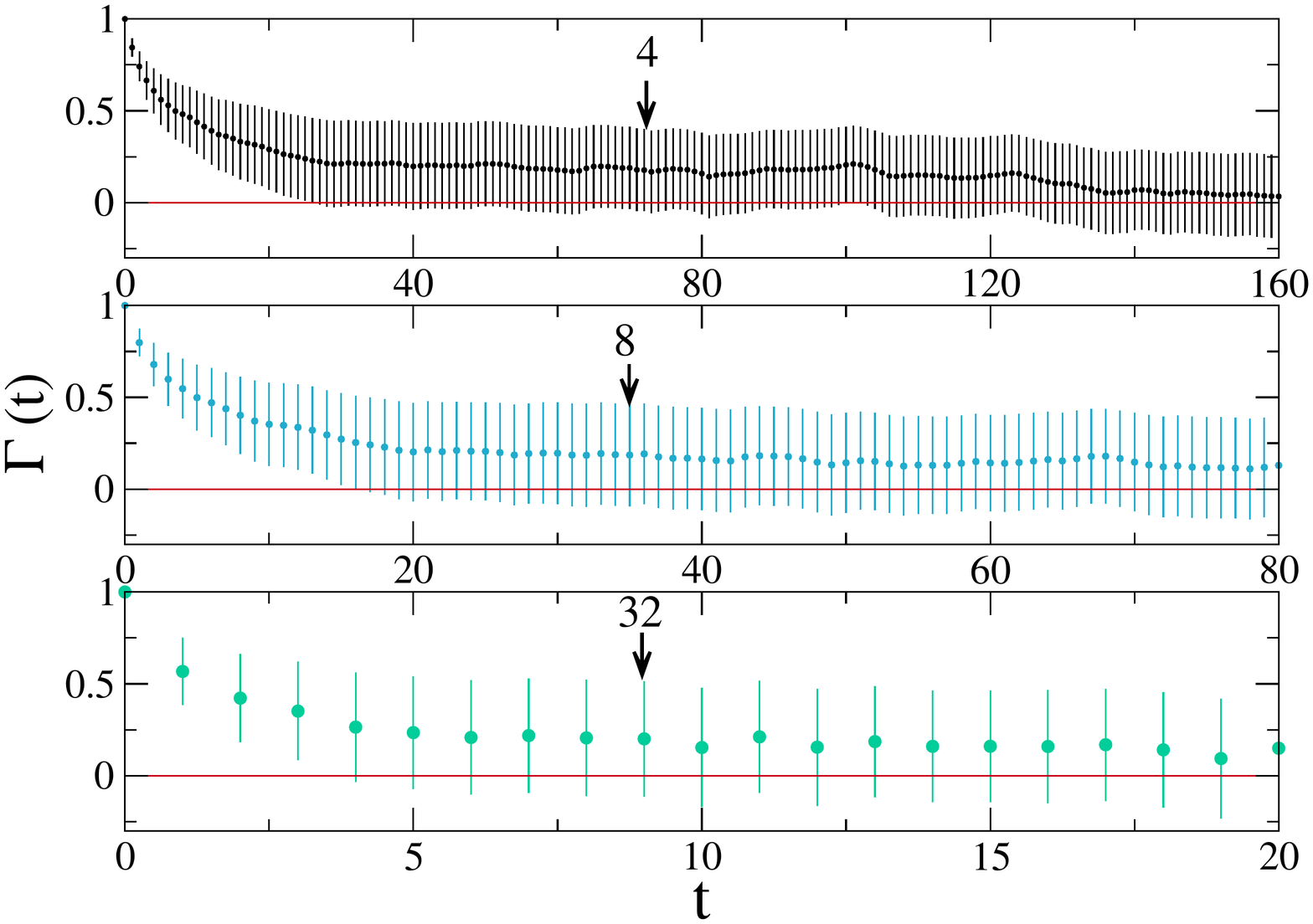}}
\subfigure{
\includegraphics[width=2.8in,clip]
{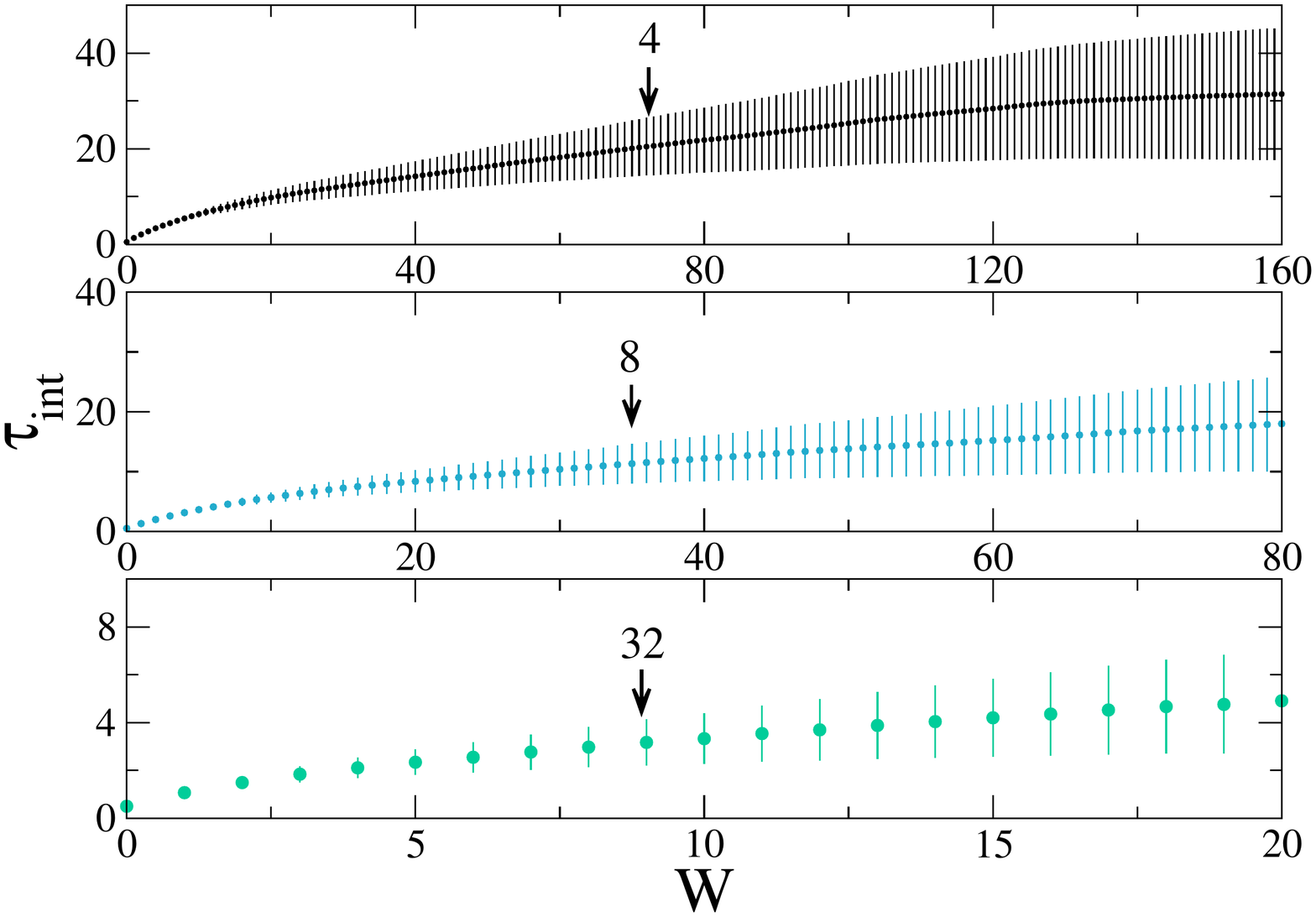}}

\caption{Normalized autocorrelation functions (left) and integrated 
autocorrelation times (right) for unsmeared plaquette with three 
different gaps ($4,8,32$) between measurements for the ensemble $C_5$ at $\beta = 5.6$. }
\label{fig_gap_plaq}
\end{figure}

Since it is exorbitant to measure 
smeared Wilson loops, propagators and smeared topological charge on each 
and every trajectory, we have measured these observables for the 
configurations saved with specific gaps. However unsmeared plaquette ($P_0$)
%and $N_{gcr}$ (GCR iteration count of the outer algorithm for   
%global correction force) are 
is measured on each trajectory. It is mandatory, however, to check that the measured 
autocorrelation scales appropriately with the gaps, so as to ensure the correct
determination of the autocorrelation. We have carried out such checks and a
typical result is presented in Figs. \ref{fig_gap_plaq}. In Fig. \ref{fig_gap_plaq} we present the    
 normalized autocorrelation functions (left) and integrated 
autocorrelation times (right) for unsmeared plaquette with three 
different gaps ($4,8,32$) between measurements for the ensemble $C_5$. 
The data clearly exhibit the scaling properties with the gaps.

%\subsection{Topological charge, susceptibility and topological charge 
% correlator}
%\begin{figure}
%\subfigure{
%\includegraphics[width=2.8in,clip]
%{kappa-dependence-integrated-autocorr-time-sus-b58.pdf}}
%\subfigure{\includegraphics[width=2.8in,clip]
%{autocorr_length_sus_kappa_158_158125.pdf}
%}
%\caption{Integrated autocorrelation times for $Q_{20}^2$ at  
%$\beta=5.8$ (left) and  $\beta=5.6$ (right).    }
%\label{kappa-dep-sus}
%\end{figure}
\begin{figure}
\subfigure{\includegraphics[width=2.8in,clip]{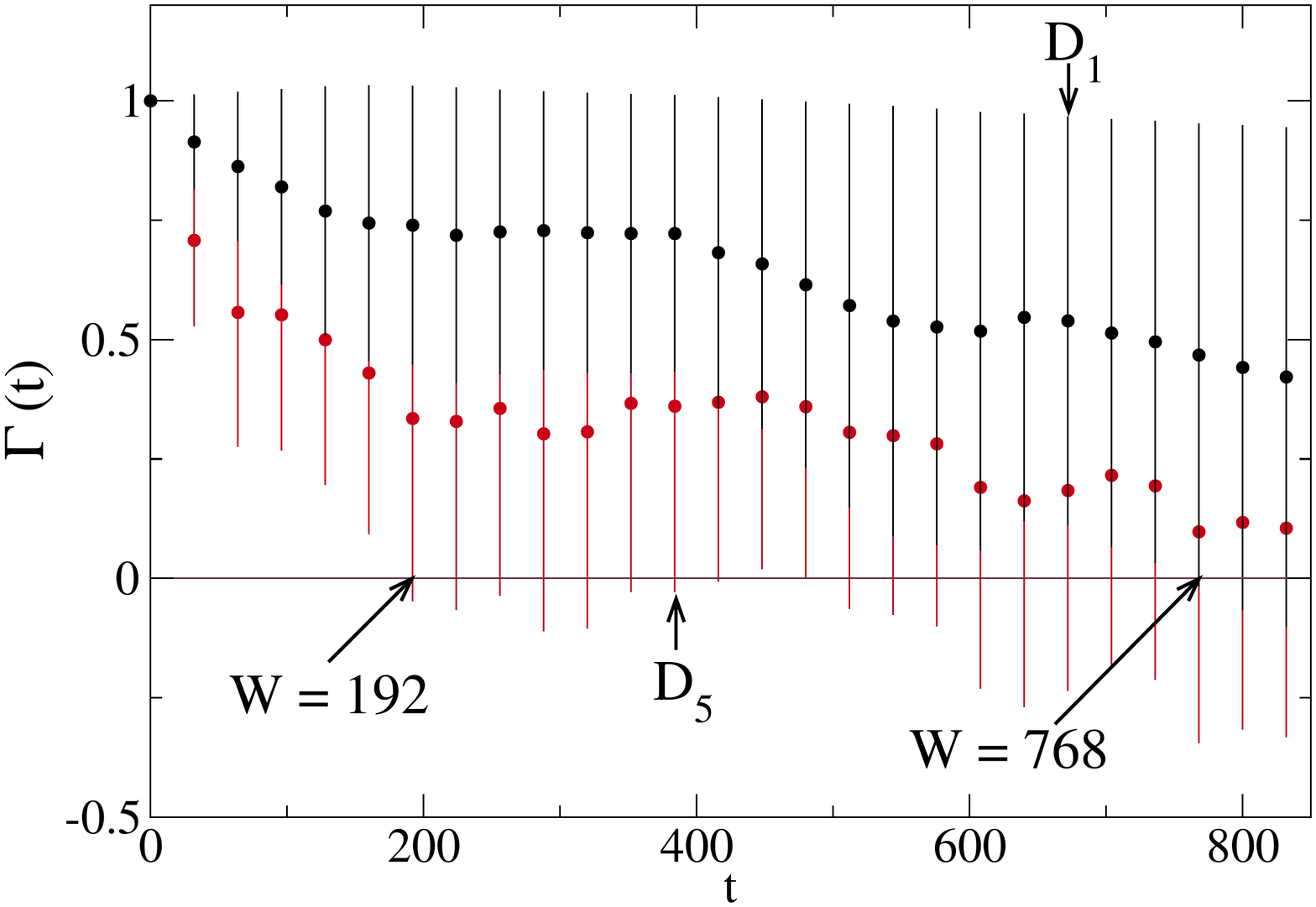}}
\subfigure{\includegraphics[width=2.8in,clip]{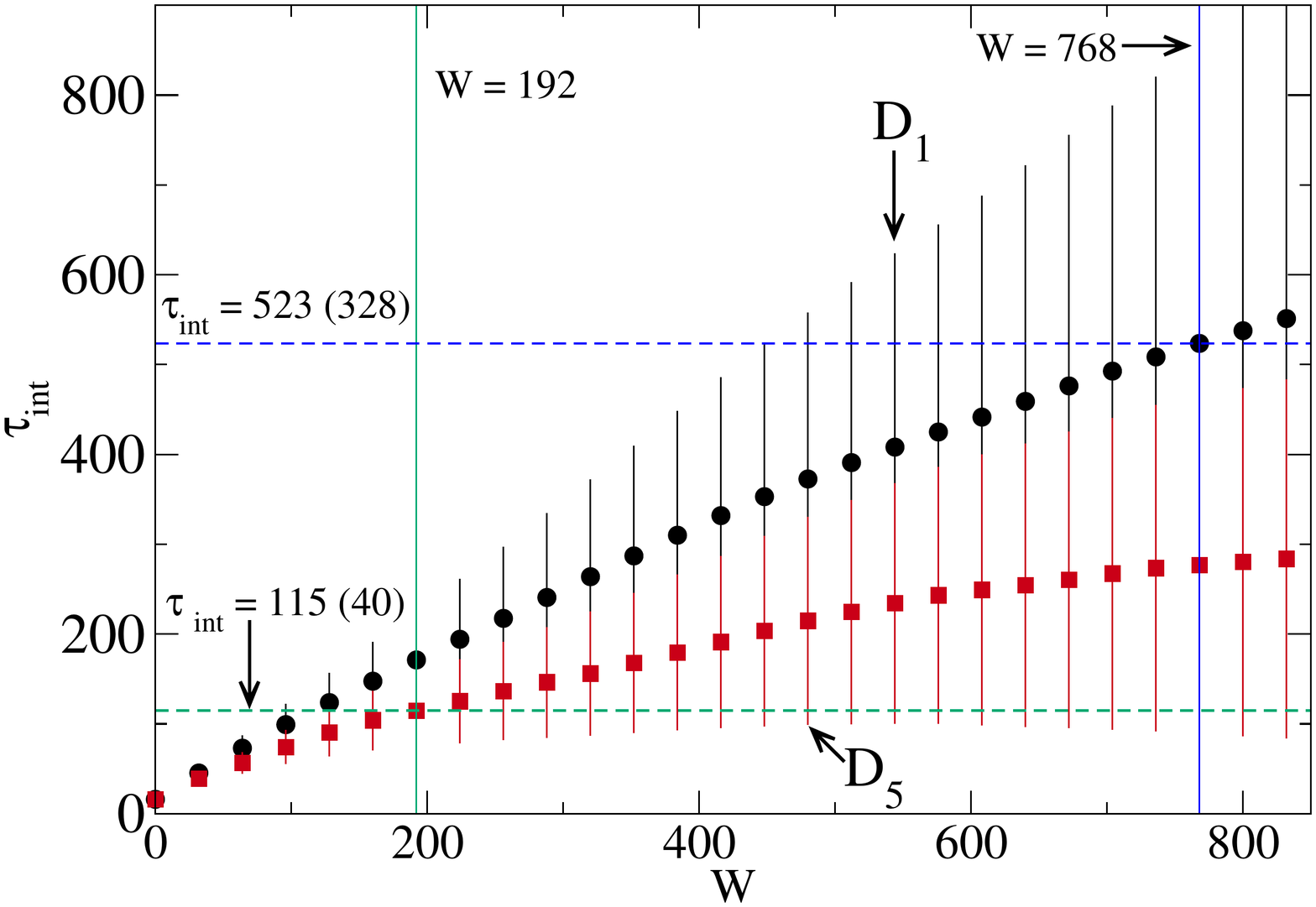}}
\caption{Normalized autocorrelation functions (left) and integrated autocorrelation times
(right) for $Q_{20}^2$
at $\beta =5.8$ for the ensembles $D_1$ and $D_5$.    }
\label{kappa-dep-sus-MS-b58}
\end{figure}
\begin{figure}

\subfigure{\includegraphics[width=3.in,clip]
{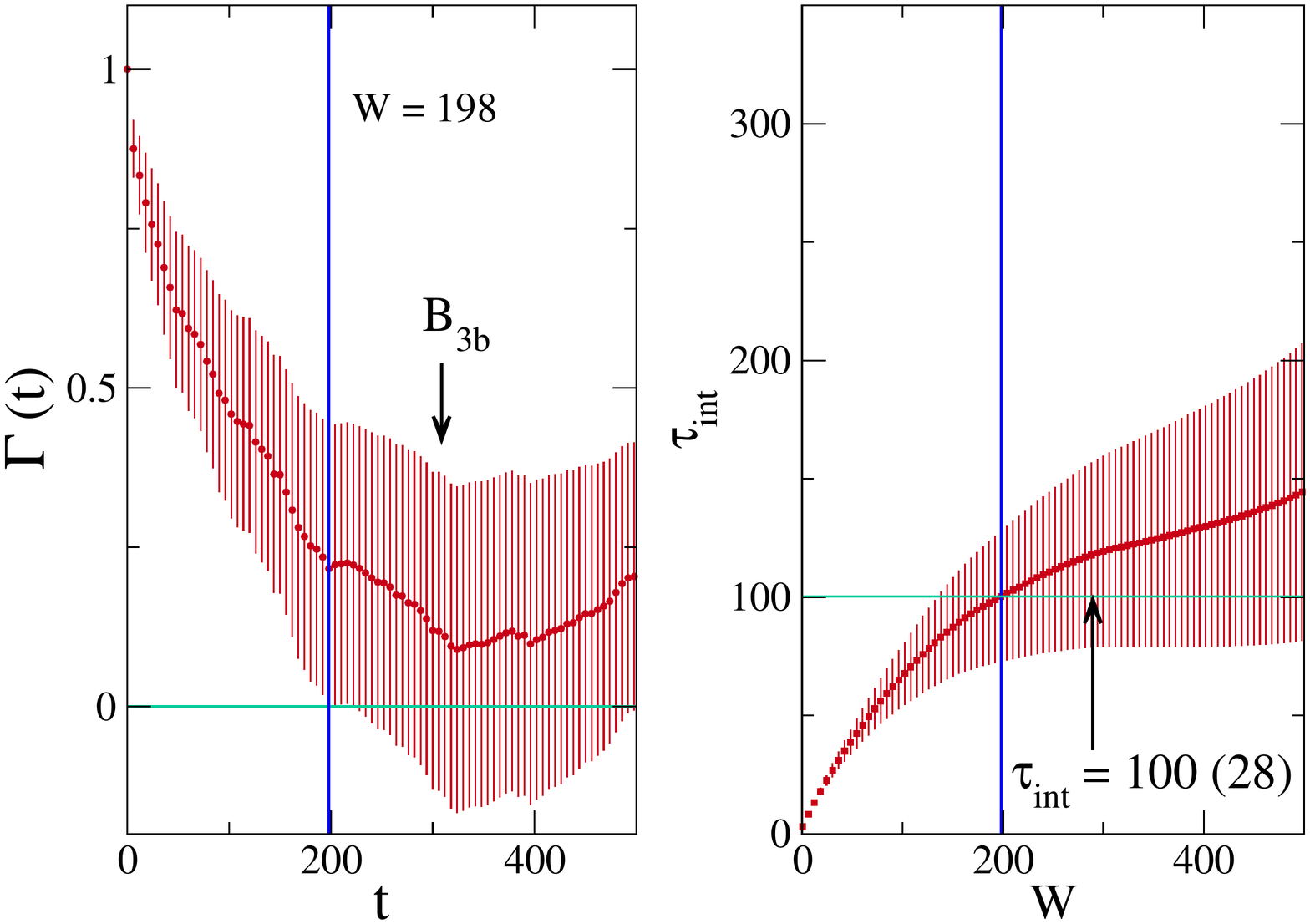}}
\subfigure{\includegraphics[width=3.in,clip]
{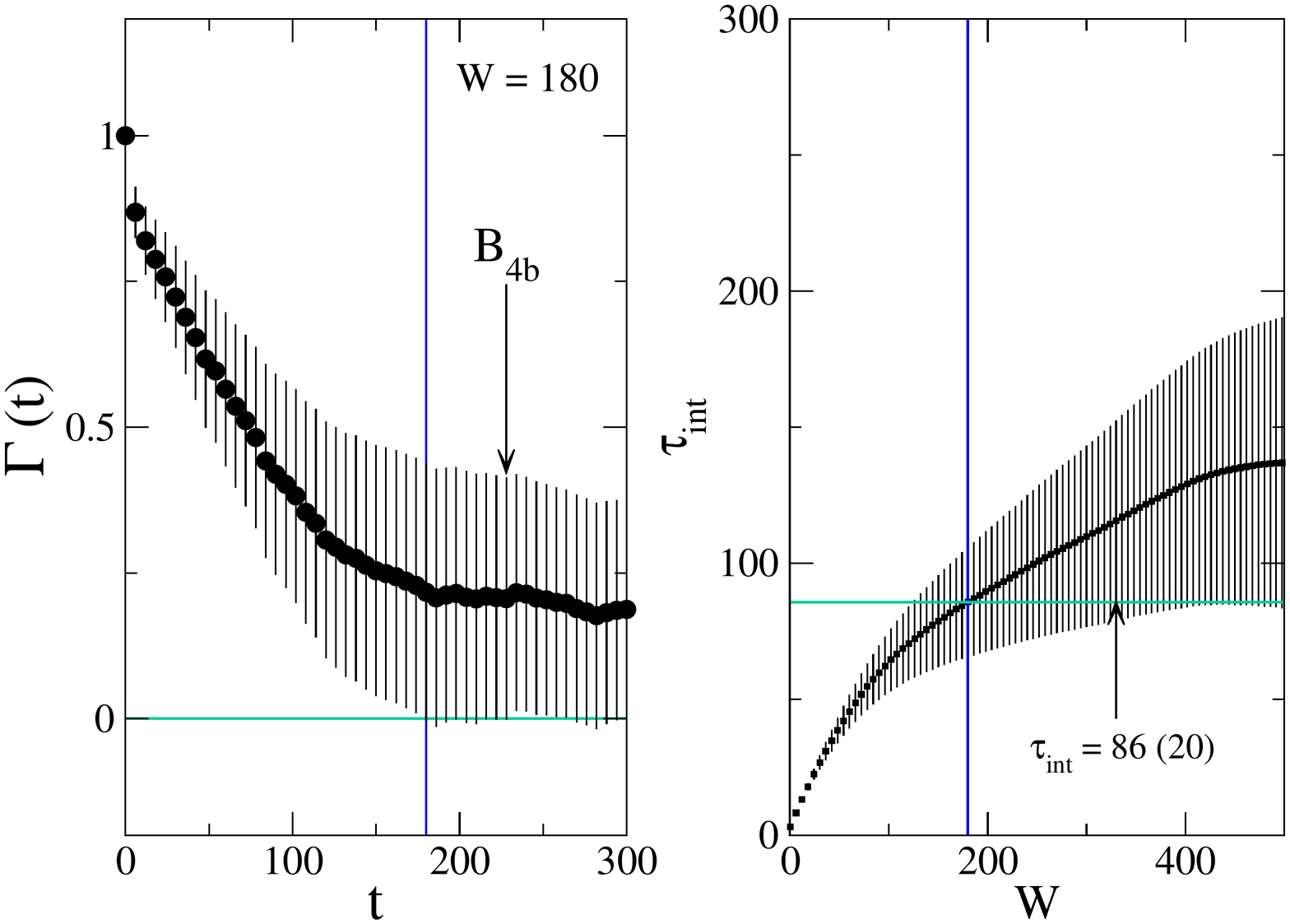}}
\begin{center}
\subfigure{\includegraphics[width=3.in,clip]
{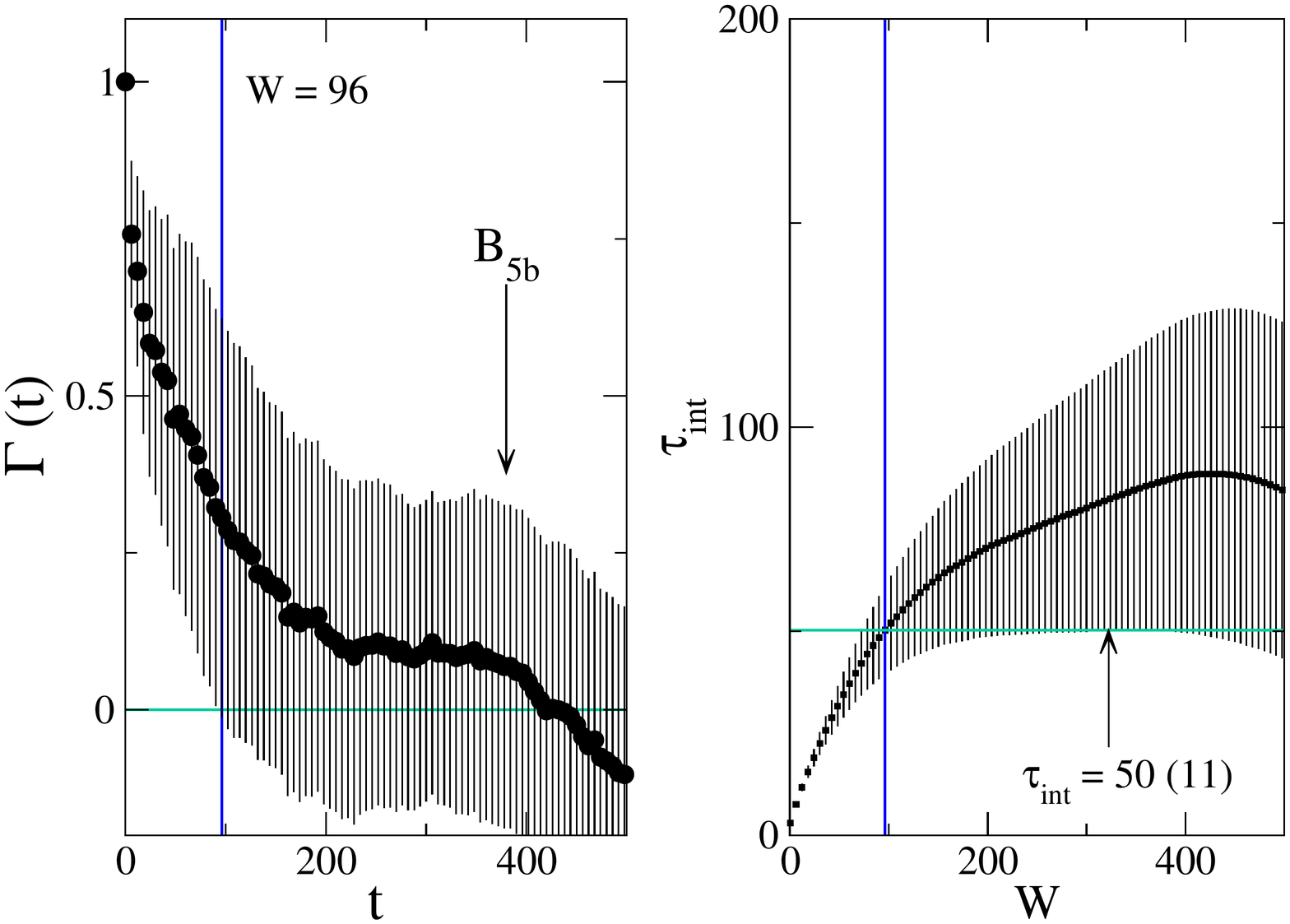}}

\caption{Normalized autocorrelation functions and integrated autocorrelation times for $Q_{20}^2$
at $\beta=5.6$ 
for the ensembles $B_{3b}$ (left), $B_{4b}$ (right) and $B_{5b}$ (bottom).    }
\label{kappa-dep-sus-MS-b56}
\end{center}
\end{figure}

%\begin{figure}
%\subfigure{\includegraphics[width=2.8in,clip]
%{beta-dependence-autocorr-func-sus.pdf}}
%\subfigure{\includegraphics[width=2.8in,clip]
%{Compare-beta-dependence-sus-correlator-kappa1543-158.pdf}}
%\caption{Normalized autocorrelation functions at $\beta=5.6$ and $\beta=5.8$
%at comparable quark masses for $Q_{20}^2$ (left),  $Q_{20}^2$ and $C(r)$ (right).    }
%\label{beta-dep-sus-corr}
%\end{figure}
\begin{figure}
\includegraphics[width=4.0in,clip]
{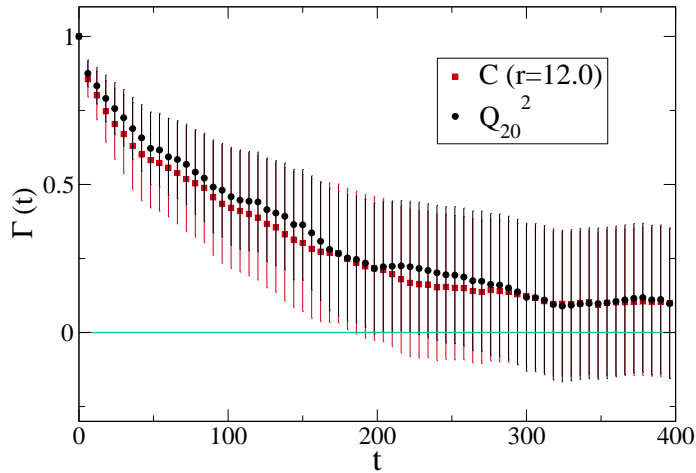}
\caption{Normalized autocorrelation functions  for $C(r=12)$ and $Q_{20}^2$ at $\beta=5.6$ for the ensemble
$B_{3b}$.    }
\label{sus-corr-b56}
\end{figure}

\begin{figure}
\subfigure{\includegraphics[width=2.8in,clip]
{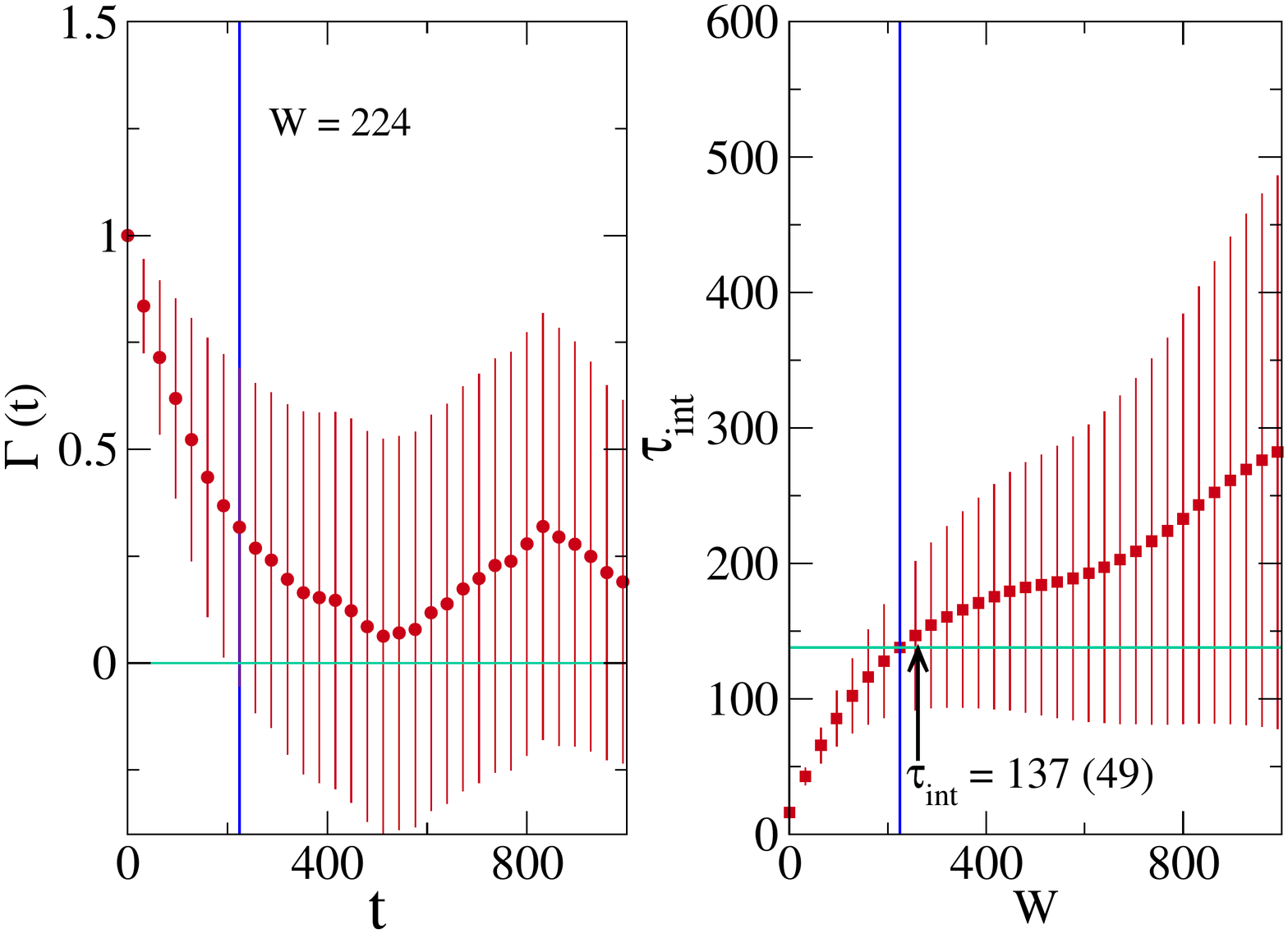}}
\subfigure{\includegraphics[width=2.8in,clip]
{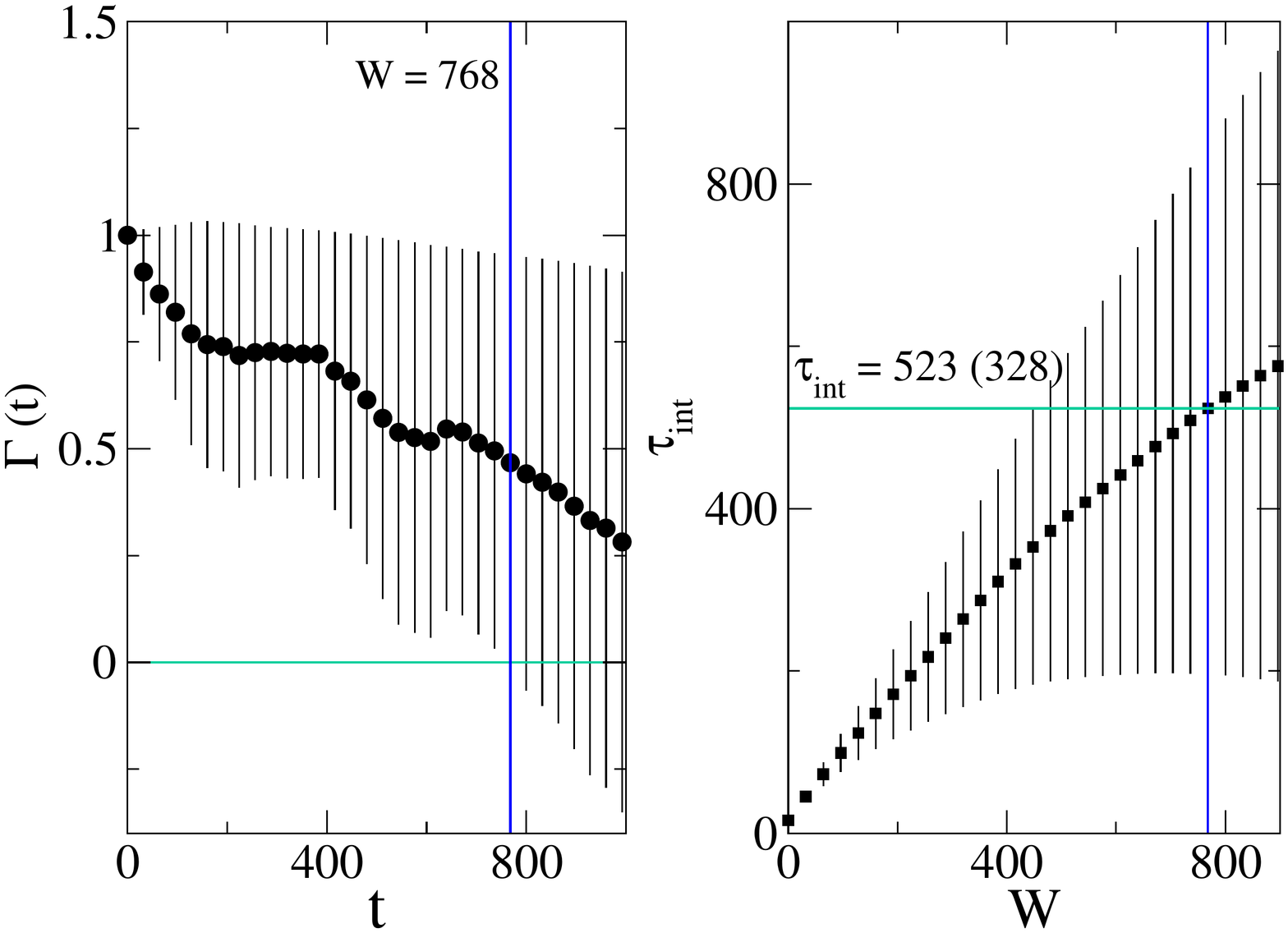}}
\caption{Normalized autocorrelation functions and integrated autocorrelation times 
for $C(r=12)$ (left) and $Q_{20}^2$ (right)
at $\beta=5.8$
for the ensemble $D_1$.    }
\label{sus-corr-b58}
\end{figure}
%In Table \ref{table3} integrated autocorrelation times for 
%topological charges ($Q_{20}$) and 
%susceptibilities ($Q^2_{20}$) after $20$ levels of HYP smearing are presented.
%Part of these data have been presented before \cite{latt2011}.
%For $D_5$ the integrated autocorrelation time plot for $Q_{20}$
%does not show clear plateau. 
 
%The integrated 
%autocorrelation times at fixed 
%molecular dynamics trajectory length ($\tau$) increase with 
%increasing $\kappa$ for both $Q_{20}$ and $Q_{20}^2$.
%Integrated autocorrelation times for both $Q_{20}$ and $Q_{20}^2$ increase
%substantially when lattice spacing is decreased.

In Figs. \ref{kappa-dep-sus-MS-b58} and \ref{kappa-dep-sus-MS-b56} we show 
noramlized autocorrelation functions and integrated autocorrelation times for $Q_{20}^2$ 
at different 
$\kappa$'s for $\beta = 5.8$ and $\beta = 5.6$ respectively.
Windows are chosen as indicated by the vertical lines.
Figs. \ref{kappa-dep-sus-MS-b58} and \ref{kappa-dep-sus-MS-b56} show that at both the lattice spacings 
($\beta =5.6, 5.8$) 
autocorrelations of $Q_{20}^2$ decrease with decreasing
quark mass even though for the smaller quark mass at $\beta =5.8$ ($D_5$)
molecular dynamics trajectory length ($\tau$) is smaller.
Note that the trend is more visible at smaller lattice spacing ($\beta =5.8$). 
A possible explanation \footnote{Stefan Schaefer (private communication).}
for this suppression of autocorrelation with decreasing quark 
mass is that the algorithm needs to span between lesser number of topological sectors at smaller
quark mass since the width of the Gaussian distribution of topological charge decreases with decreasing
quark mass. 

In Fig. \ref{sus-corr-b56} we show 
normalized autocorrelation functions  for $C(r=12)$ and $Q_{20}^2$ at $\beta=5.6$ for the ensemble
$B_{3b}$. 
Fig. \ref{sus-corr-b56} shows that at $\beta = 5.6$ the autocorrelations for 
$Q_{20}^2$ and $C(r=12)$ are very close.
In Fig. \ref{sus-corr-b58} we show
normalized autocorrelation functions and integrated autocorrelation times 
for $C(r=12)$ (left) and $Q_{20}^2$ (right)
at $\beta=5.8$
for the ensemble $D_1$ where pion mass is comparable with 
the pion mass for the ensemble $B_{3b}$.
Fig. \ref{sus-corr-b58} shows that at $\beta = 5.8$ the autocorrelation for
$Q_{20}^2$ is larger than the autocorrelation for $C(r=12)$. 
Figs. \ref{sus-corr-b56} and \ref{sus-corr-b58} show that autocorrelation for $Q_{20}^2$
increases quite significantly with decreasing lattice spacing at comparable
pion mass whereas the autocorrelation of topological 
charge density correlator ($C(r)$) 
increases slightly with 
decreasing lattice spacing. Taking into account the effect of active link ratio (see for example section 3.1 in Ref. \cite{sommer})
($R=0.363$ for $\beta = 5.6$ and $R = 0.422$ for $\beta = 5.8$) strengthens this conclusion. 

\begin{figure}
\subfigure{\includegraphics[width=2.8in,clip]{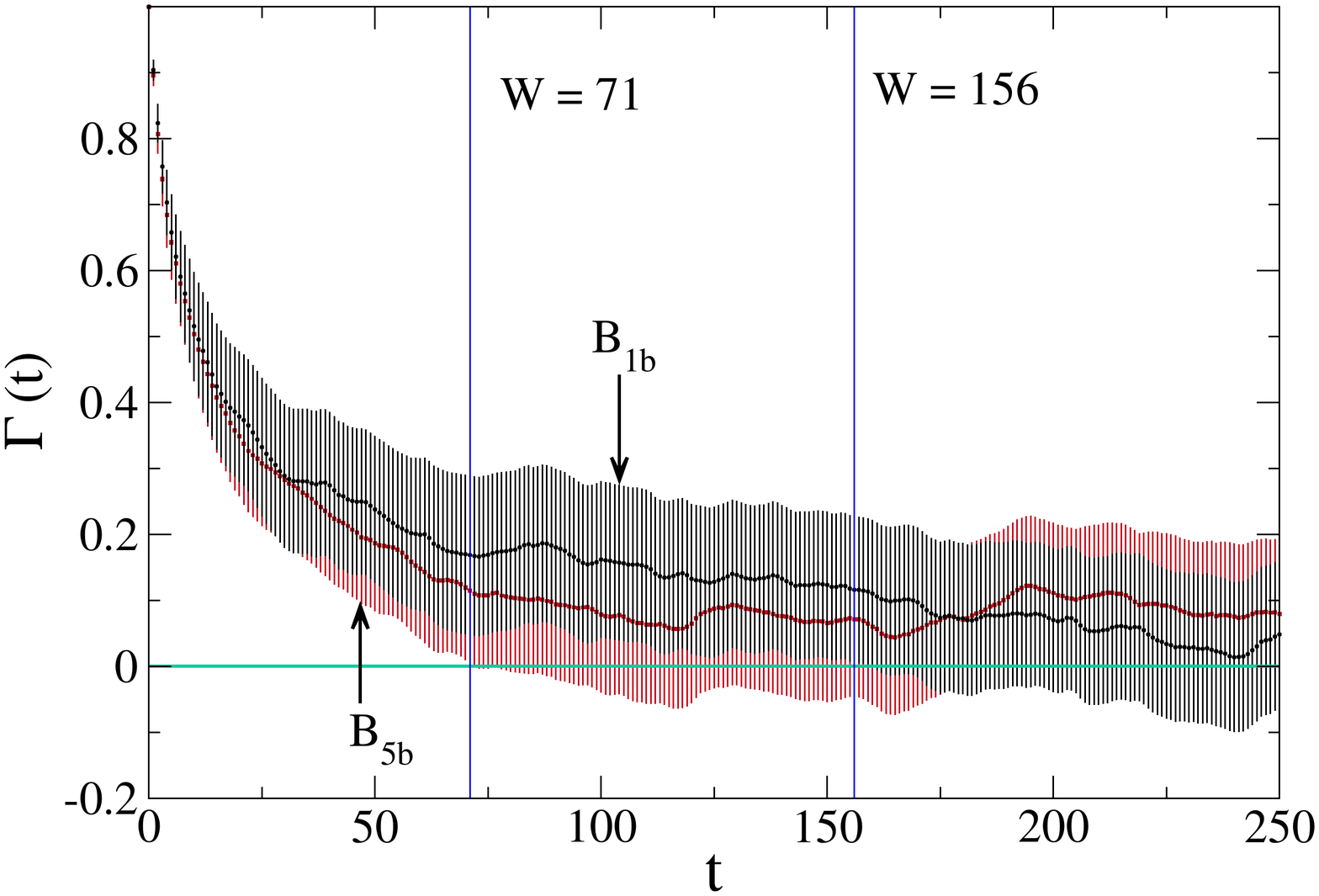}}
\subfigure{\includegraphics[width=2.8in,clip]{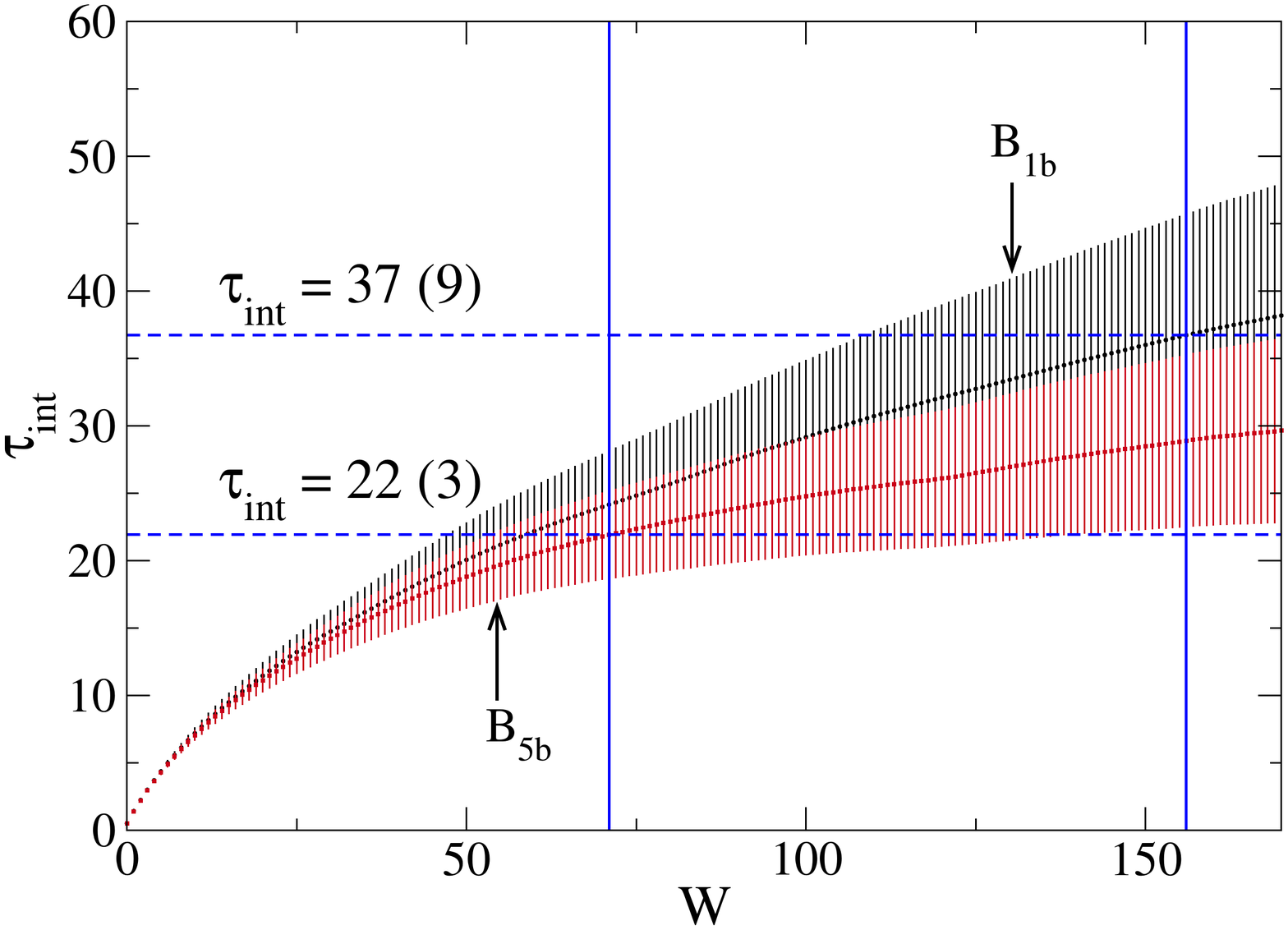}}
\caption{Normalized autocorrelation functions (left) and integrated autocorrelation times
(right) for $P_{0}$
at $\beta =5.6$ for the ensembles $B_{1b}$ and $B_{5b}$.    }
\label{kappa-dep-P0}
\end{figure}

\begin{figure}
\subfigure{\includegraphics[width=2.8in,clip]{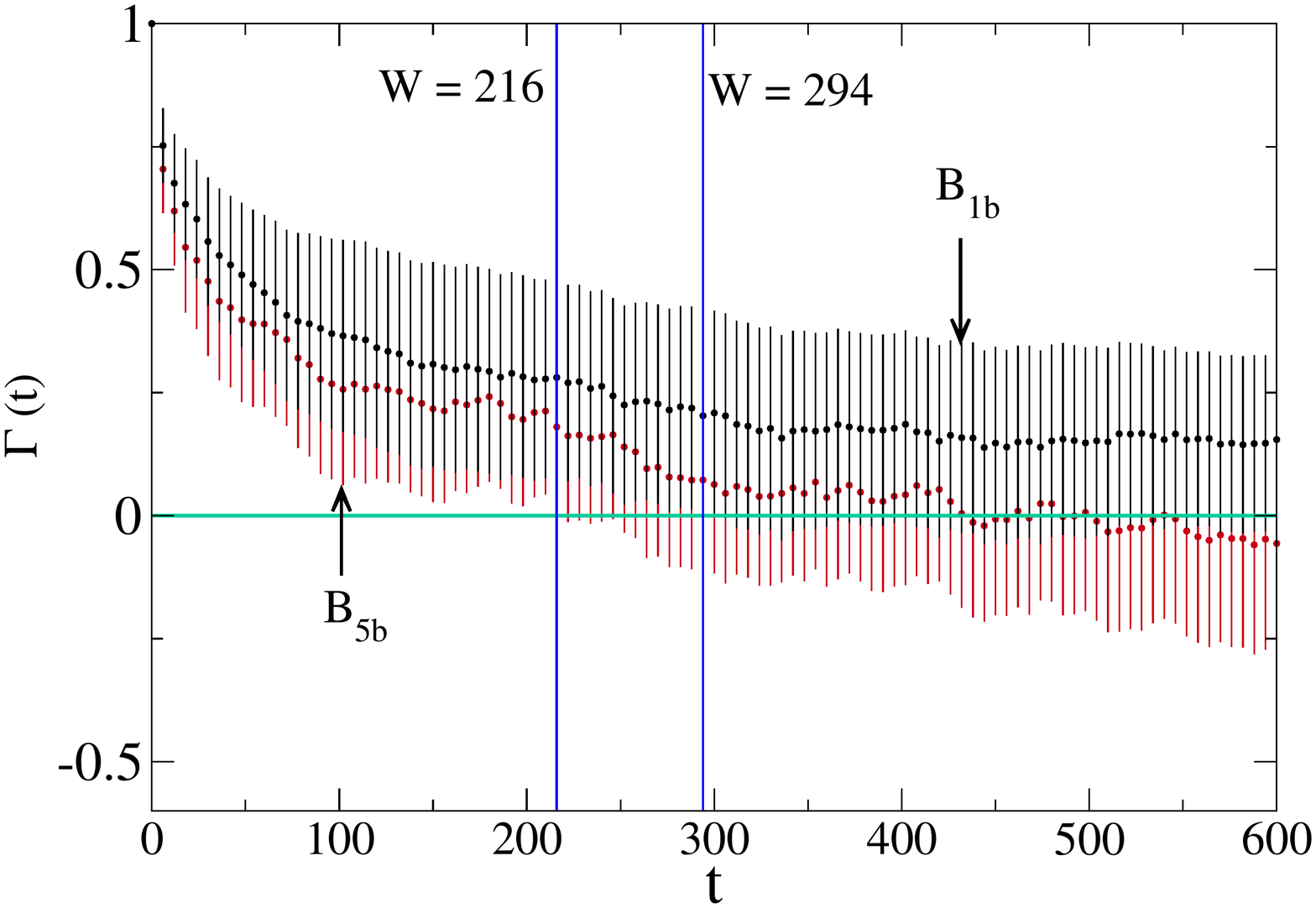}}
\subfigure{\includegraphics[width=2.8in,clip]{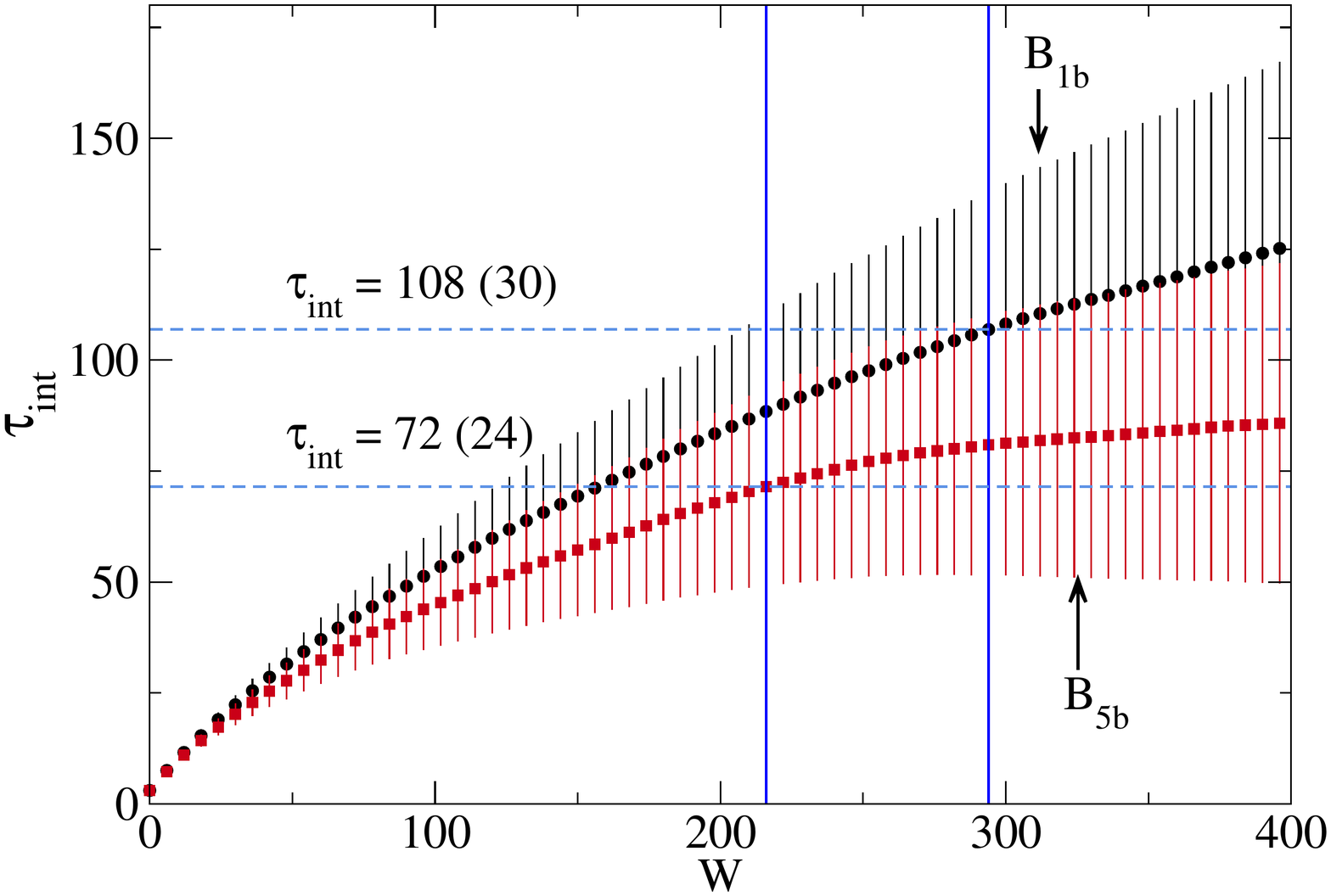}}
\caption{Normalized autocorrelation functions (left) and integrated autocorrelation times
(right) for $W_{20}(4,4)$
at $\beta =5.6$ for the ensembles $B_{1b}$ and $B_{5b}$.    }
\label{kappa-dep-W20}
\end{figure}

%******************************************************************************************%
%\subsection{Plaquette and Wilson loop}
%In Table \ref{table2} integrated autocorrelation times for 
% $P_0$ and 
%$P_{20}$ at $\beta = 5.6$ and lattice volume $32^3\times 64$ are presented.
%The data show decrase of autocorrelation with decreasing quark mass for both the quantities.
In Fig. \ref{kappa-dep-P0}
we present normalized autocorrelation functions (left)
and integrated autocorrelation times (right)
for $P_0$ for  $\beta = 5.6$ at lattice volume 
$24^3\times 48$.
We find that $\tau_{int}$ for $P_0$
is not increasing with decreasing quark mass.
%The figure (left panel) shows no significant quark mass dependence
%of autocorrelation for $P_0$, even though comparison of autocorrelations
%between largest and smallest quark masses indicates that there is some decreasing 
%trend of autocorrelation with decreasing quark mass. Right panel
%shows that autocorrelation of $W_{20}(4,4)$ decreases with decreasing quark mass.
%\subsection{Dependence on Lattice Spacing}
%Fig. \ref{beta-dep-P0} shows that $P_0$ decorrelates faster in smaller lattice
%spacing at comparable quark mass. 
%Fig. \ref{beta-dep-P20-W20} shows faster decorrelation of $P_{20}$ (left)
%and $W_{20}(4,4)$ (right) 
%in smaller lattice spacing at comparable quark mass.
In Fig. \ref{kappa-dep-W20} 
we present normalized autocorrelation functions (left)
and integrated autocorrelation times (right)
for $W_{20}(4,4)$ for  $\beta = 5.6$ at lattice volume 
$24^3\times 48$. 
The figure shows  $\tau_{int}$ for $W_{20}(4,4)$ 
is also not increasing with decreasing quark mass.

%\begin{figure}
%\includegraphics[width=4.0in,clip]{beta-dependence-p_0-kappa-158-1543.pdf}
%\caption{Normalized autocorrelation functions for $P_0$ at $\beta =5.6$ and $\beta =5.8$ at comparable quark masses.}
%\label{beta-dep-P0}
%\end{figure}

%\begin{figure}
%\subfigure{\includegraphics[width=2.8in,clip]{beta-dependence-autocorr-func-P20-b56-58.pdf}}
%\subfigure{\includegraphics[width=2.8in,clip]{beta-dependence-wloop-44-autocorr-func.pdf}}
%\caption{Normalized autocorrelation functions at $\beta = 5.6$ and $\beta = 5.8$ in comparable quark masses
%for $P_{20}$ (left) and $W_{20}(4,4)$ (right).    }
%\label{beta-dep-P20-W20}
%\end{figure}

\begin{figure}
\subfigure{
\includegraphics[width=3.0in,clip]{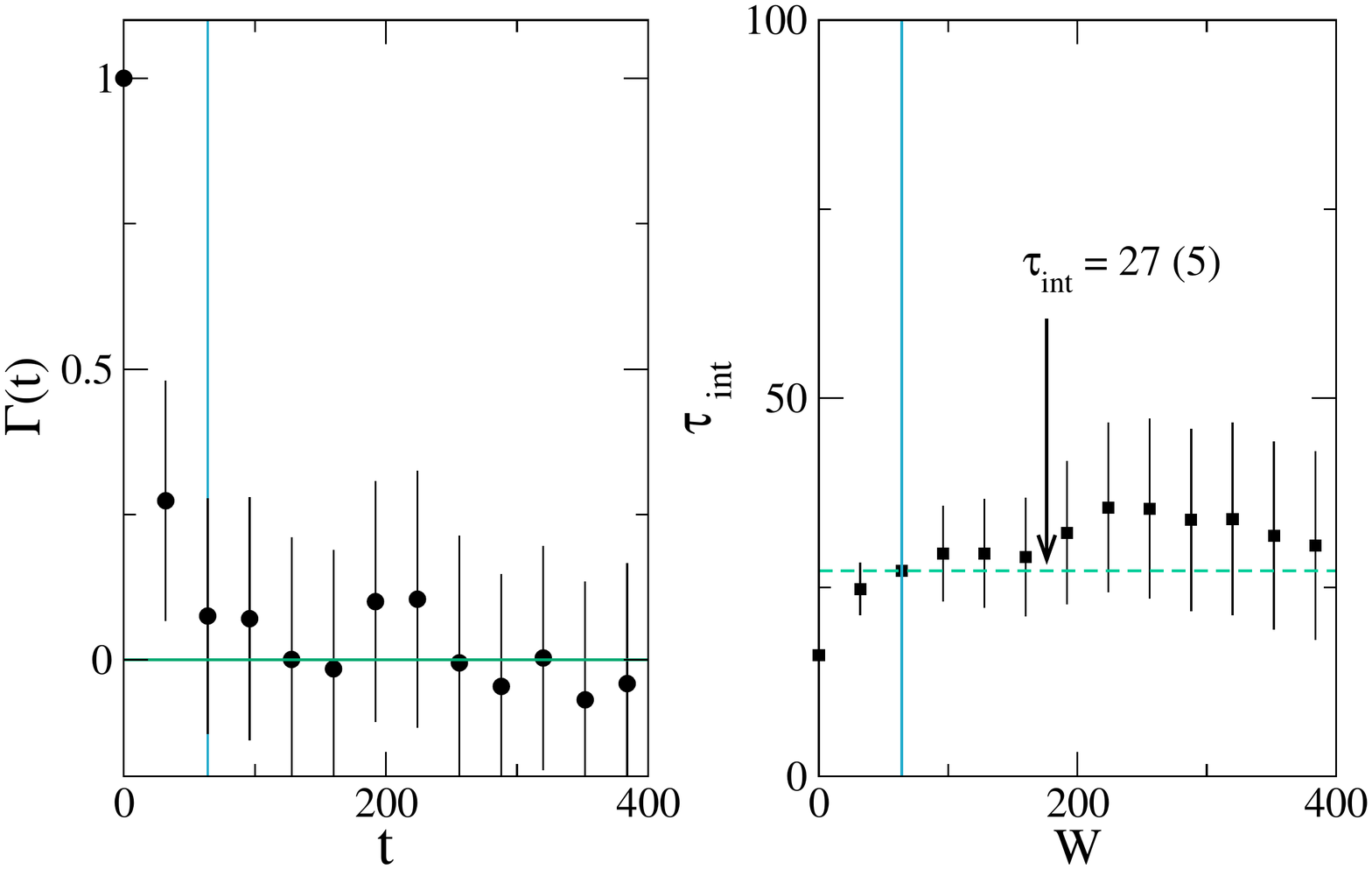}}
\subfigure{\includegraphics[width=3.0in,clip]{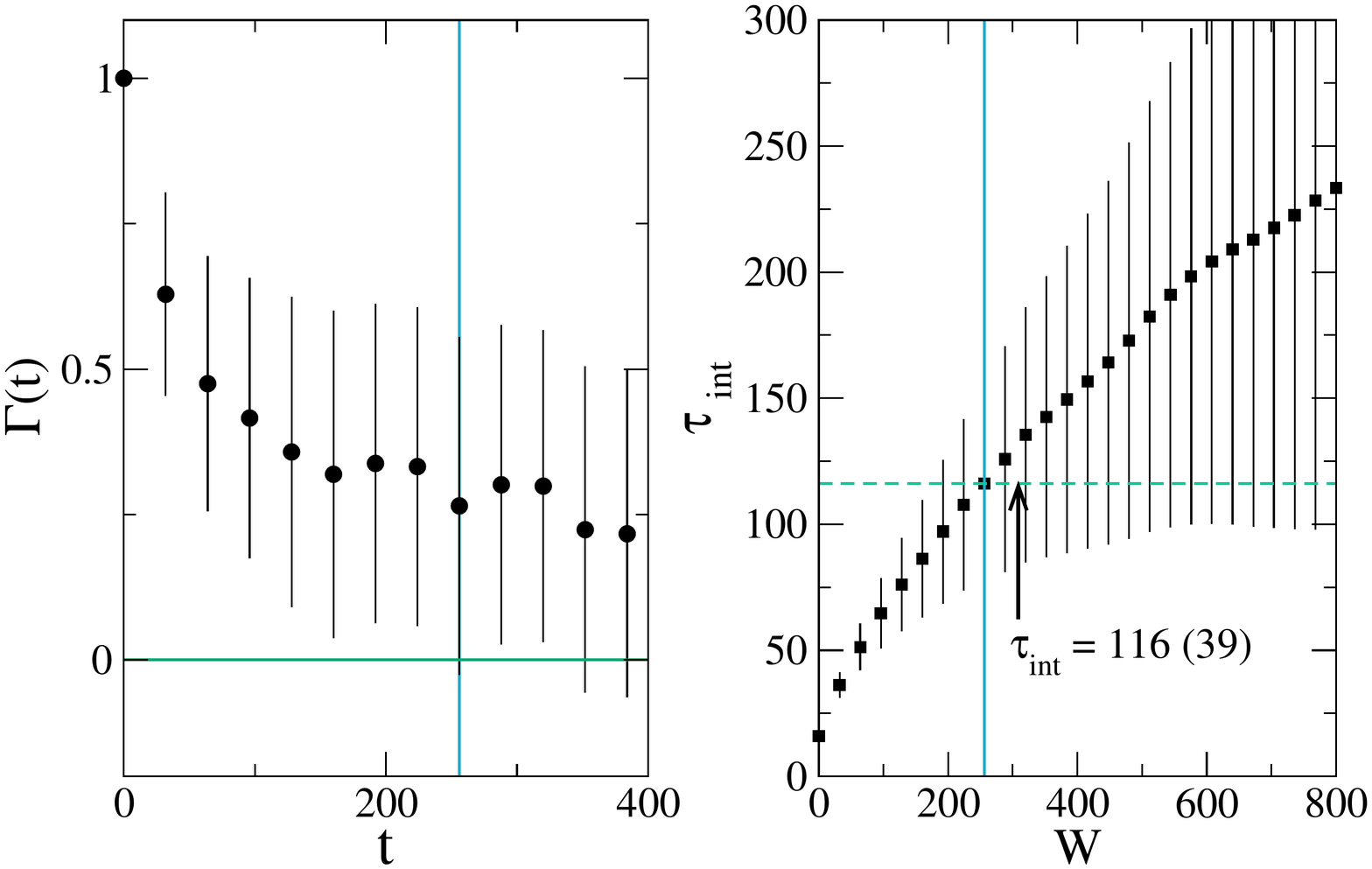}
}
\caption{Normalized autocorrelation functions and 
integrated autocorrelation times for Wilson loops with 
$R=1$, $T=1$ (left) and $R=4$, $T=5$ (right) at $\beta = 5.8$ for the ensemble $D_1$.    }
\label{wloop_size}
\end{figure}

\begin{figure}
\subfigure{
\includegraphics[width=3.0in,clip]{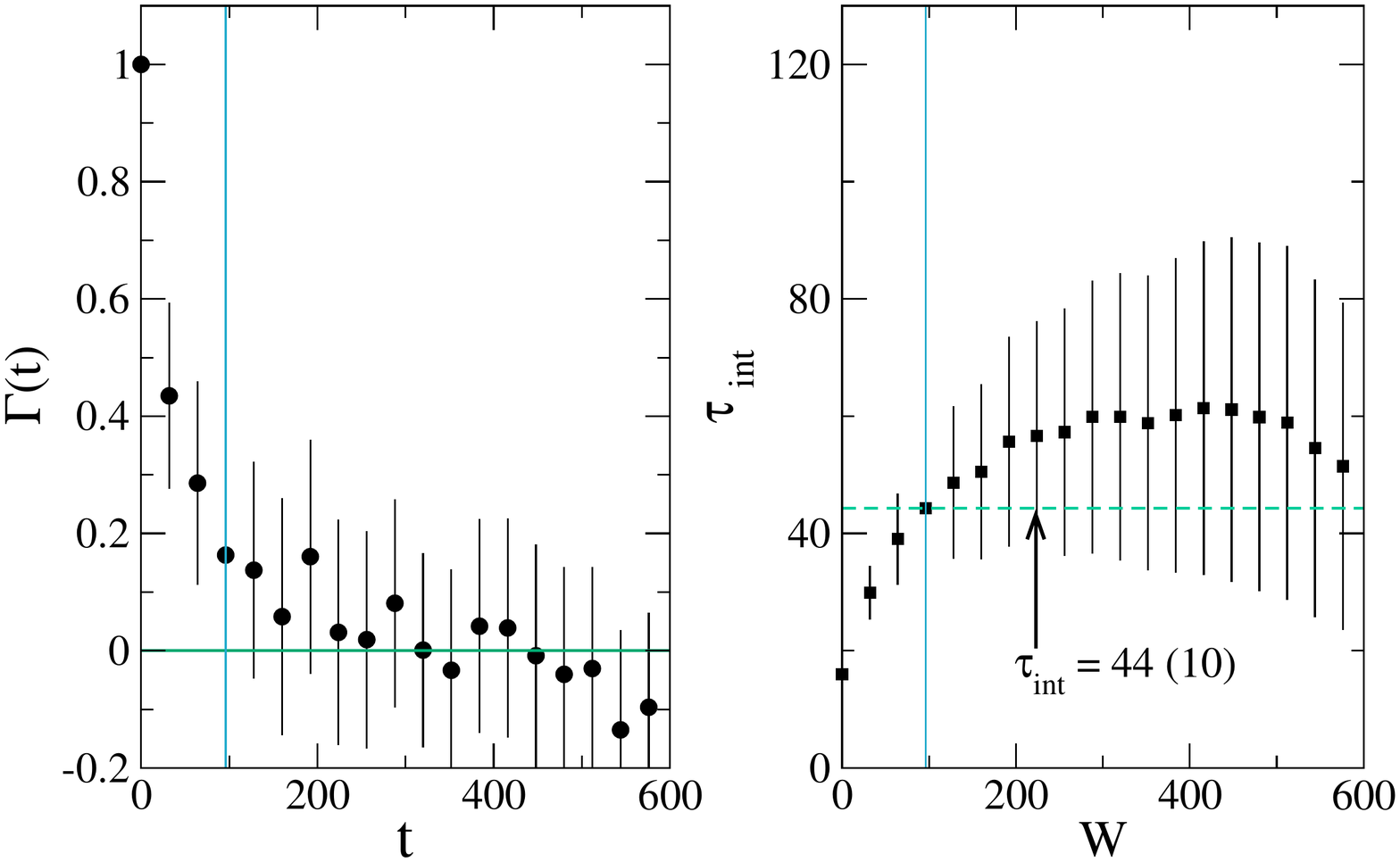}}
\subfigure{\includegraphics[width=3.0in,clip]{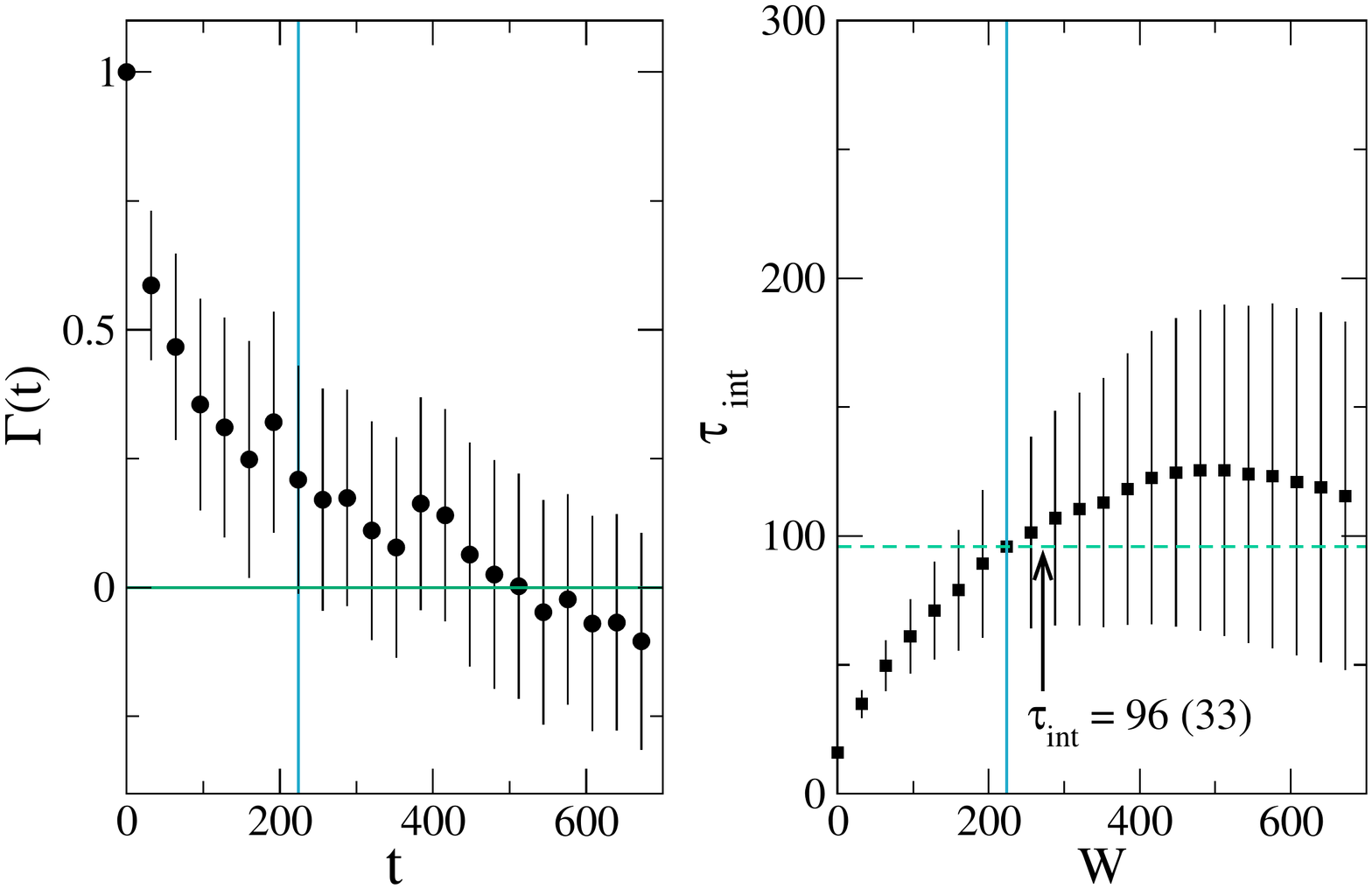}
}
\caption{Normalized autocorrelation functions and 
integrated autocorrelation times for Wilson loops with 
smearing levels $=5$ (left) and 
smearing levels $=40$ (right) at $\beta = 5.6$ for the ensemble $C_2$.    }
\label{wloop_smlev}
\end{figure}

%*****************************************************************************************************%
%\subsection{Effect of smearing}
For the measurement of static potential $V(r)$ one needs to 
measure Wilson loops
of various sizes. In the measurement 
of a Wilson loop, to suppress 
unwanted fluctuations smearing is needed. 
Therefore it is interesting to study how autocorrelation of Wilson loops 
changes with sizes of the Wilson loops 
and smearing levels. In Fig. \ref{wloop_size} we present normalized autocorrelation functions  
and integrated autocorrelation times for $W_{20}$ with different
sizes for the ensemble $D_1$.
In Fig. \ref{wloop_smlev} we show normalized autocorrelation functions
and integrated autocorrelation times for $W(3,3)$ with different 
levels of HYP smearing for the ensemble $C_2$.
We observe that the integrated autocorrelation time increases with
the increasing size of the Wilson loop and also with the 
increasing smearing level.
In the context of Wilson loop and Polyakov loop, SESAM collaboration has observed
that geometrically extended observables suffer more from autocorrelation \cite{sesam}
with HMC algorithm. 


\begin{figure}
\includegraphics[width=4in,clip]{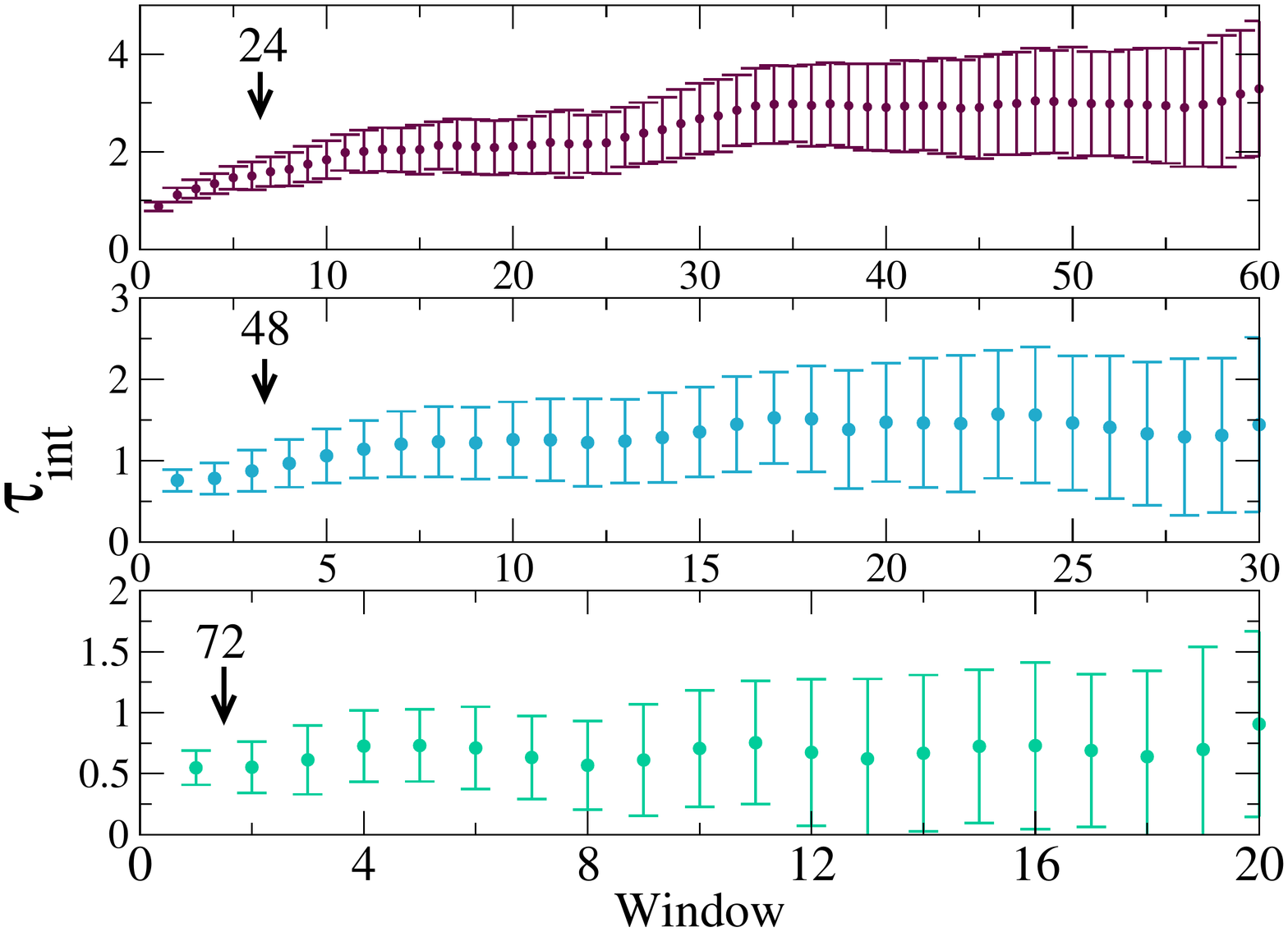}
\caption{Integrated 
autocorrelation times for $PP$ propagator with wall source for three 
different gaps ($24,48,72$) between measurements for the ensemble $B_{4a}$ at $\beta = 5.6$.}
\label{fig_gap_prop}
\end{figure}

For hadronic observables the autocorrelations are quite small and since the number of 
measurements are not large the errors calculated from Eqs. (\ref{fn-err}) and (\ref{tau-err}) are quite large
and mask the trends of the central values of autocorrelations. 
Our emphasis is on different trends of autocorrelations. To detect some trend 
of the central values of the autocorrelations for the hadronic observables 
we use a rough estimate of errors 
by single omission jackknife technique.    
In Fig. \ref{fig_gap_prop} integrated 
autocorrelation times for $PP$ propagator with wall source for three 
different gaps ($24,48,72$) between measurements for the ensembles $B_{4a}$
are presented. The data clearly exhibit the scaling properties with the gaps.
In Table \ref{table4} integrated autocorrelation times for pion ($PP$) and
nucleon propagators 
with wall sources
at a given time slice
are presented. 
Clearly the integrated autocorrelation time decreases with 
increasing $\kappa$ (i.e. decreasing quark mass)
both for pion and nucleon propagators. Similar observation was made by
ALPHA collaboration in the case of Clover fermion \cite{marinkovic}.
The autocorrelation times of
pion and nucleon propagators with point source and sink (not presented here) 
are smaller than
the gap with which configurations are saved.

\begin{figure}
\includegraphics[width=4in,clip]{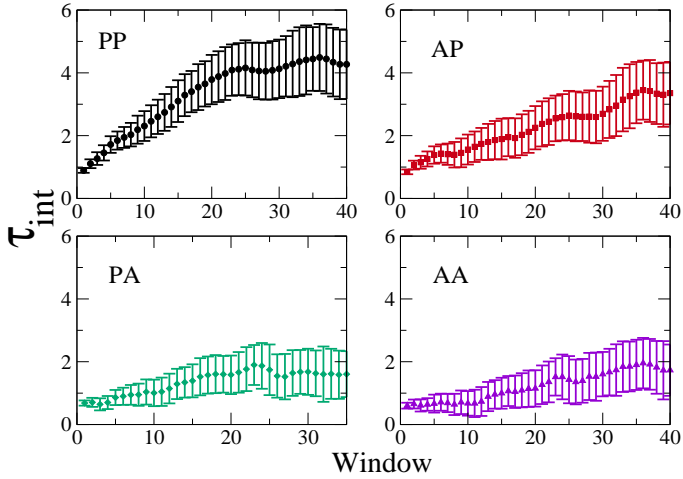}
\caption{Integrated autocorrelation times for $PP$, $AP$, $PA$ and
$AA$ correctors with wall source for the ensemble $B_{3a}$ at $\beta = 5.6$. Measurements are done with a gap of 24 trajectories.  }
\label{fig_PP_AP_PA_AA}
\end{figure}
\begin{table}
\begin{center}
\begin{tabular}{|l|l|l|}

 \multicolumn{3}{c}{$\beta = 5.6$} \\

\hline
$tag$&$\tau_{int}^{Pion}$& $\tau_{int}^{Nucleon}$\\ 
\hline
%{$16^3\times32$}&{$0.156$}&{HMC}&{5000}&{200}&{0.5}&{16(6)}&{28(10)(200)}&{21(7)}&{10(5)(200)}&{} \\
%{$~~~~~,,$}&{$0.157$}&{HMC}&{5000}&{200}&{0.5}&{23(8)}&{27(6)(200)}&{32(8)}&{15(5)(200)}&{} \\
%{$~~~~~,,$}&{$0.1575$}&{HMC}&{5000}&{200}&{0.5}&{26(7)}&{26(7)(200)}&{34(9)}&{16(5)(200)}&{} \\ 
%{$~~~~~,,$}&{$0.158$}&{HMC}&{5000}&{200}&{0.5}&{25(9)}&{36(10)(200)}&{83(20)}&{29(9)(200)}&{} \\ 
%{$~~~~~,,$}&{$0.15775$}&{$~~~~8^4$}&{16320}&{510}&{0.5}&{119(38)}&{68(18)(255)}&{440(84)}&{276(68)(255)}&{} \\ 
%{$~~~~~,,$}&{$~~~,,$}&{$8^3\times16$}&{9408}&{294}&{0.5}&{37(10)}&{36(8)(294)}&{254(39)}&{70(17)(294)}&{} \\
%\hline 
%**********************************************************************
\hline
%{$B_{1a}$}&{$23(7)$}&{$14(3)$}\\
%{$B_{2a}$}&{$132(32)$}&{$72(24)$}\\
%{$A_3$}&{$$}&{$~~~,,$}&{$12^3\times6$}&{4992}&{0.5}\\
{$B_{3a}$}&{$99(19)$}&{$75(18)$}\\
{$B_{4a}$}&{$50(9)$}&{$34(9)$}\\
%{$~~~~~,,$}&{$0.158125$}&{$6^3\times8$}&{11808}&{492}&{0.5}&{62(18)}&{122(35)(123)}&{635(69)}&{327(42)(492)}&{}\\
{$B_{5a}$}&{$40(10)$}&{$25(9)$}\\
%{$~~~~~,,$}&{$0.15825$}&{$6^3\times8$}&{9360}&{390}&{0.5}&{()}&{()(90)}&{238(30)}&{161(22)(390)}&{107(18)} \\
%{$~~~~~,,$}&{$0.15825$}&{$6^3\times8$}&{8640}&{360}&{0.5}&{62(18)}&{133(50)}&{()}&{()}&{}\\
%**********************************************************************
\hline
%{$32^3\times64$}&{$0.15775$}&{$8^3\times16$}&{6080}&{190}&{0.5}&{86(40)}&{443(93)(94)}&{723(134)}&{416(152)(94)}&{}\\
{$C_2$}&{$39(13)$}&{$33(17)$}\\
%{$~~~~~,,$}&{$0.15815$}&{$8^3\times16$}&{11520}&{360}&{0.5}&{115(41)}&{210(84)(90)}&{}&{()}&{}\\
{$C_3$}&{$31(15)$}&{$26(7)$}\\
{$C_4$}&{$34(11)$}&{$18(6)$}\\
%{$B_4$}&{$155(35)$}&{$55(21)$}\\
%{$~~~~~,,$}&{$0.1584$}&{$8^3\times16$}&{3904}&{122}&{0.25}&{37(25)}&{25(18)}&{338(112)}&{127(54)}&{} \\
%{$B_5$}&{$28(20)$}&{$40(12)$}\\
\hline \hline

%***********************************************************************
%***********************************************************************
 % \multicolumn{4}{c}{$\beta = 5.8$} \\
%\hline
%$tag$&$\tau_{int}^{Ngcr\_frf}$& $\tau_{int}^{Plaq.}$& $\tau_{int}^{Wloop}$\\
%\hline
%{$32^3\times64$}&{$0.154$}&{$8^3\times16$}&{6400}&{200}&{0.5}&{20(8)}&{(200)}&{74(23)}&{28(18)(rej=500)(200)}&{} \\
%{$~~~~~,,$}&{$0.1541$}&{$8^3\times16$}&{6400}&{200}&{0.5}&{69(27)}&{(100)}&{}&{(100)}&{} \\
%{$C_1$}&{$41(14)$}&{$29(7)$}&{$265(44)$}\\
%{$C_2$}&{$53(8)$}&{$37(9)$}&{$222(39)$}\\
%{$C_3$}&{$75(22)$}&{$48(23)$}&{$356(69)$}\\
%{$C_4$}&{$100(15)$}&{$53(8)$}&{$218(52)$}\\
%{$C_5$}&{$53(11)$}&{$48(14)$}&{$421(67)$}\\
%\hline \hline
\end{tabular}
\end{center}
%\end{center}
\caption{Integrated autocorrelation times for pion ($PP$) and
nucleon propagators with wall sources at $\beta = 5.6$.}
\label{table4}
\end{table}

For the determination of pion decay constants and PCAC quark mass, pion 
propagators other than $PP$ are also needed. 
In Fig. \ref{fig_PP_AP_PA_AA} the integrated autocorrelation times for $PP$, $AP$, $PA$ and $AA$ correlators with wall source for the ensemble $B_{3a}$
are presented.
The propagators with $A$ in the source are less correlated than 
$P$ in the source.

\begin{figure}
\subfigure{
\includegraphics[width=2.8in,clip]
{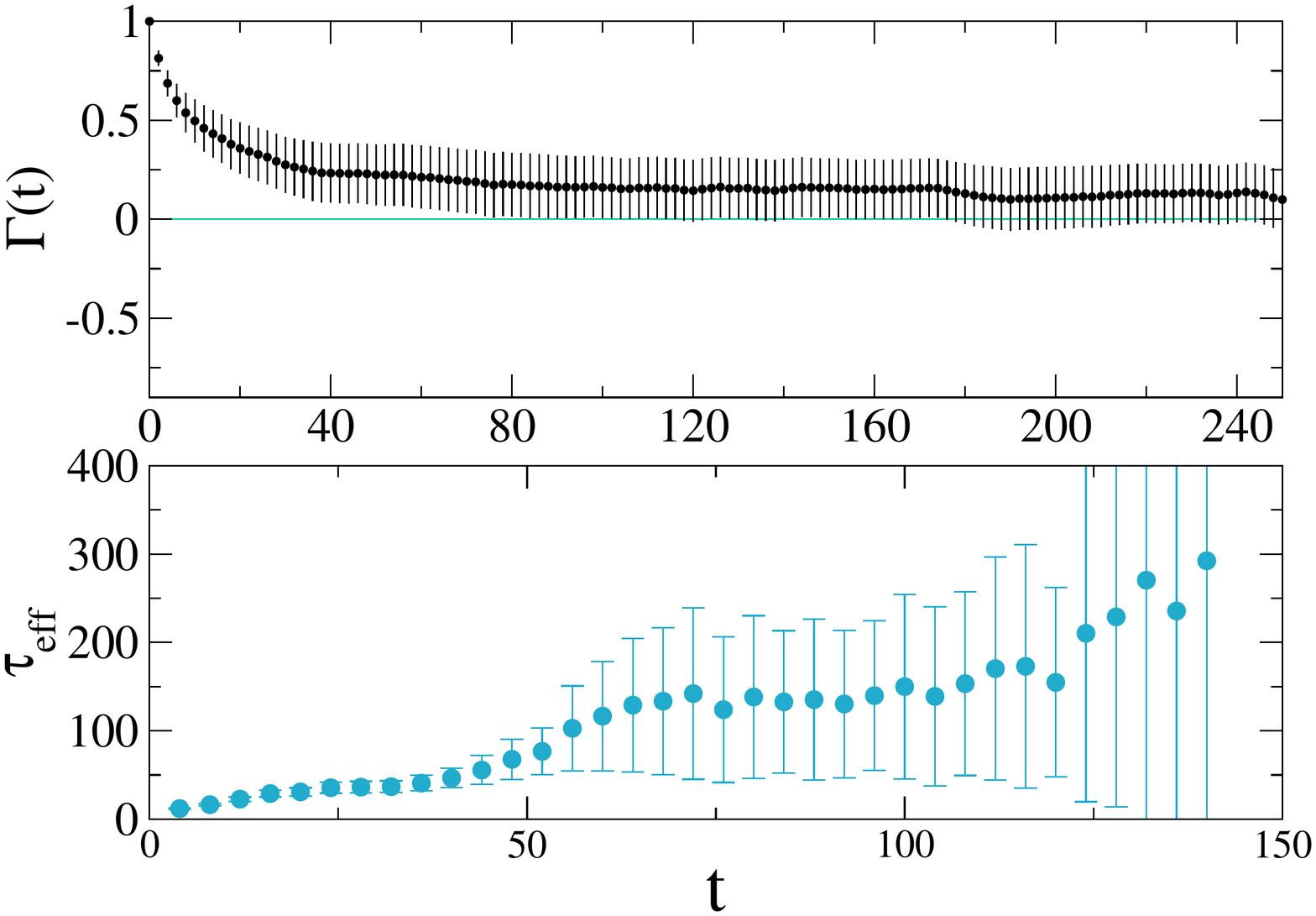}}
\subfigure{\includegraphics[width=2.8in,clip]
{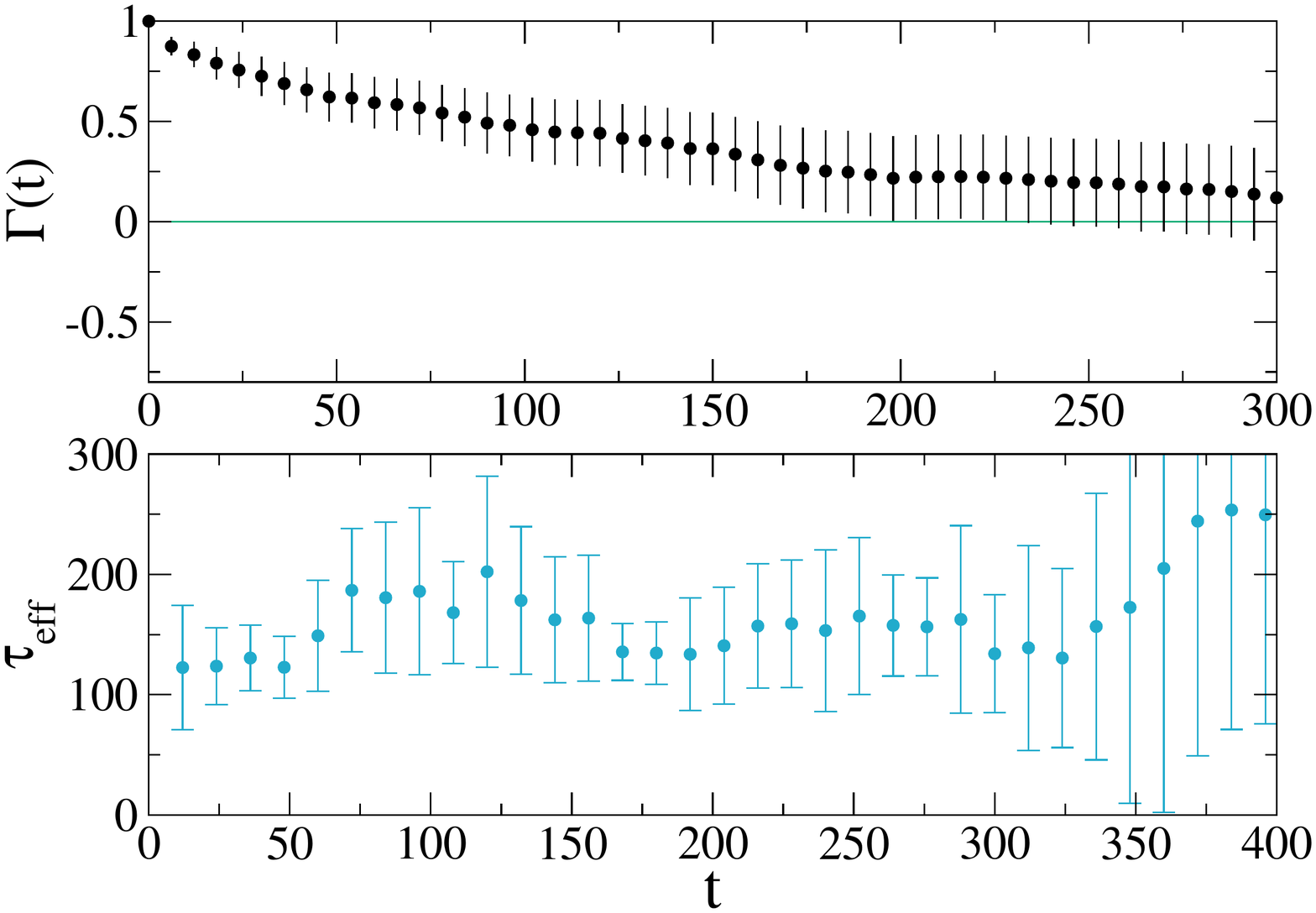}}
\caption{Normalized autocorrelation function and effective autocorrelation
time for $P_0$ (left) $Q^2_{20}$ (right) for the ensemble $B_{3b}$ at $\beta = 5.6$.}
\label{effective_tau_exp}
\end{figure}
%%%%%%%%%%%%%%%%%%%%%%%%%%%%%%%%%%%%%%%%%%%%%%%%%%%%%%%%%%%%%%%%%%%%%%%%%%%%%%%
\section{Improved estimation of $\tau_{int}$}
%%%%%%%%%%%%%%%%%%%%%%%%%%%%%%%%%%%%%%%%%%%%%%%%%%%%%%%%%%%%%%%%%%%%%%%%%%%%%%%

We have seen that the autocorrelations of  
different observables behave differently with the change in lattice spacing.
As pointed out in \cite{sommer}, this behaviour is controlled by the coupling
of different observables with the slow modes of the transition matrix 
associated with Monte Carlo Markov chain. In this reference authors have  
proposed a method to quantify this coupling 
and estimate $\tau_{int}$ more 
reliably.
Following Ref. \cite{sommer}, an improved estimation of $\tau_{\rm int}$
can be determined as follows.
Let $\tau^*$ be the best estimate of the dominant time constant. If for an
observable ${\cal O}$ all relevant time scales are smaller or of the same
order of $\tau^*$   then the upper bound of  $\tau_{\rm int}$
\be
\tau^{u}_{\rm int} ~=~ \frac{1}{2}~ + \Sigma_{t=1}^{W_u} \Gamma^{\cal O}(t)
+ A_{\cal O}(W_u)~ \tau^* \label{tau-u}
\ee
where $A_{\cal O} = {\rm max}(\Gamma^{\cal O}(W_u), 2 \delta 
\Gamma^{\cal O}(W_u))$. $W_u$ is chosen where the 
autocorrelation is still significant.
One possible estimation of $\tau^*$ is by measuring effective autocorrelation time, which is
introduced in Ref. \cite{sommer} as described below.
Define effective exponential autocorrelation time 
\begin{eqnarray}
\tau_{eff}^{exp}({\cal O})= \frac{t}{2 
\ln{\frac{\Gamma^{\cal O}(t/2)}{\Gamma^{\cal O}(t)}}}.
\end{eqnarray} 
$\tau_{eff}^{exp}$ which can be an estimate of $\tau^*$ is defined as,
\begin{eqnarray} 
\tau_{eff}^{exp}= Max_{\mathcal O}\left[\frac{t}{2 \ln{
\frac{\Gamma^{\cal O}(t/2)}{\Gamma^{\cal O}(t)}}}\right].
\end{eqnarray}
The estimation of $\tau_{eff}^{exp}({\cal O})$ requires good signal to noise ratio in 
the asymptotic region in a case by case basis which in turn requires very 
long Markov chain and is beyond
the scope of the present work.

\begin{figure}
\subfigure{
\includegraphics[width=2.8in,clip]
{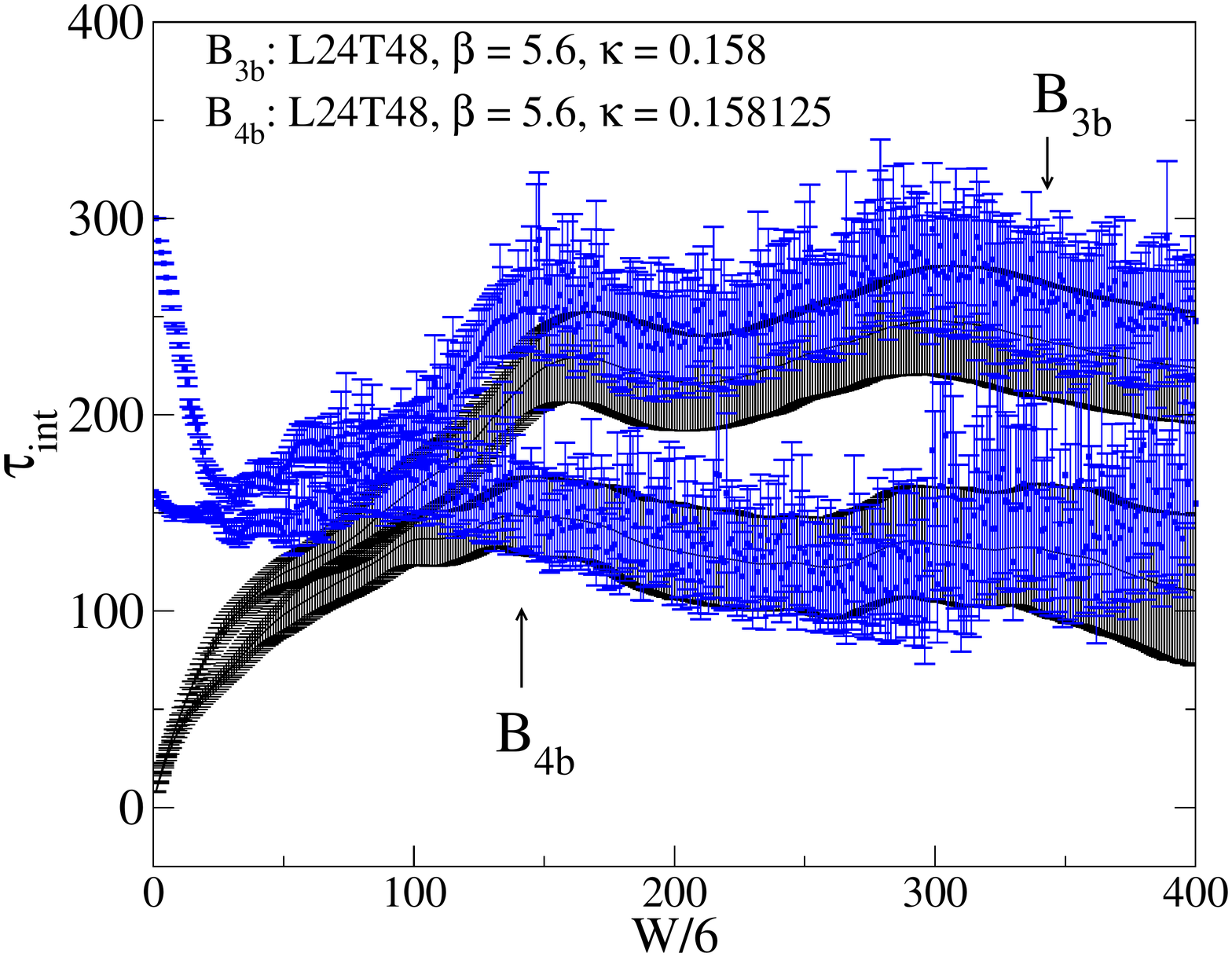}}
\subfigure{\includegraphics[width=2.8in,clip]
{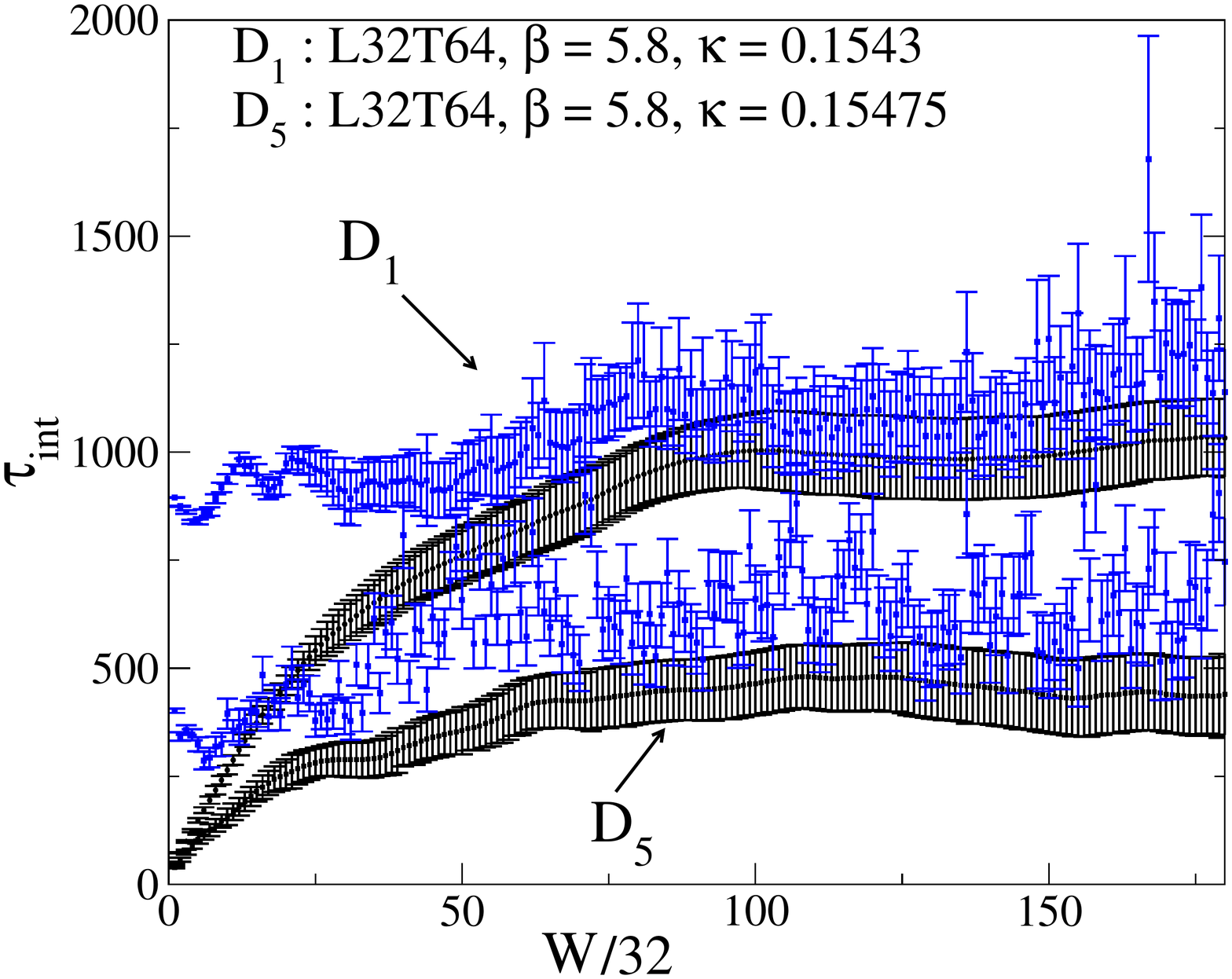}}
\caption{Integrated autocorrelation times and their upper bounds 
($\tau^u_{int}$) 
for
topological susceptibilities ($Q^2_{20}$) at $\beta$ = 5.6 
(left) and at
$\beta$ = 5.8 (right).}
\label{effective_tau}
\end{figure}

 However it is interesting to look at 
 $\tau_{eff}^{exp}({\cal O})$ where reliable data is available and we present such 
an example in Fig. \ref{effective_tau_exp} (the jackknife technique is used to 
calculate the error of $\tau_{eff}^{exp}({\cal O})$). 
In Fig.  \ref{effective_tau_exp}
it appears that $Q_{20}^2$ is coupling dominantly with slow mode,
whereas $P_0$ is coupling with more than one modes; nevertheless the 
slowest mode appearing in $P_0$ is approximately the same as in $Q_{20}^2$. This is reflected 
in the behaviour of $\tau_{eff}^{exp}({\cal O})$, which shows a single plateau for
$Q_{20}^2$, but for $P_0$, there is more than one plateau
and the data is more noisy. Similar behaviour
is observed in pure gauge theory in Ref. \cite{sommer}.

In improved estimation given in Eq. (\ref{tau-u}) central value of 
$\tau_{int}$ gets modified. To check if this modification preserves the trend of 
autocorrelation of $Q_{20}^2$ with respect to quark mass, 
in Fig. \ref{effective_tau} we present the 
integrated autocorrelation times and their upper bounds 
($\tau^u_{int}$) with rough errors estimated by jackknife method for
topological susceptibilities ($Q^2_{20}$) at $\beta$ = 5.6 
(left) and at $\beta$ = 5.8 (right).
 At both 
lattice spacings, we find that both $\tau_{int}(Q^2_{20})$ and
$\tau^u_{int}(Q^2_{20})$ decrease as quark mass decreases.

In conclusion, an extensive study of autocorrelation of several 
observables in lattice QCD with two degenerate 
flavours of naive Wilson fermion has shown that 
(1) at a given lattice spacing, autocorrelations of 
topological susceptibility and pion and nucleon propagators with 
wall source 
 decrease with decreasing quark mass and autocorrelations of plaquette and Wilson loop
do not increase with decreasing quark mass,
(2) autocorrelation of topological susceptibility 
substantially increases with decreasing lattice spacing
but autocorrelation of topological charge density correlator shows only mild increase and
(3) increasing the size and the smearing level increase the
autocorrelation of Wilson loop.

{\bf Acknowledgements} 
\vskip .05in

We thank Stefan Schaefer for a critical reading of an earlier version of the
manuscript and for suggestions for improvement of the manuscript. 
  Numerical calculations are carried out on Cray XD1 and Cray XT5 systems 
supported 
by the 10th and 11th Five Year Plan Projects of the Theory Division, SINP under
the DAE, Govt. of India. We thank Richard Chang for the prompt maintenance of 
the systems and the help in data management. This work was in part based on 
the public lattice gauge theory codes of the 
MILC collaboration \cite{milc} and  Martin L\"{u}scher \cite{ddhmc}.

%%%%%%%%%%%%%%%%%%%%%%%%%%%%%%%%%%%%%%%%%%%%%%%%%%%%%%%%%%

%%%%%%%%%%%%%%%%%%%%%%%%%%%%%%%%%%%%%%%%%%%%%%%%%%%%%%%%%%%%%%%%%%%%%%%%   

%%%%%%%%%%%%%%%%%%%%%%%%%%%%%%%%%%%%%%%%%%%%%%%%%%%

\end{document}